\documentclass{article}
\usepackage{amsmath}

\input{tcilatex}
\begin{document}

\author{Nick Laskin\thanks{%
Email address: nlaskin@rocketmail.com}}
\title{\textbf{Time Fractional Quantum Mechanics}\\
}
\date{TopQuark Inc.\\
Toronto, ON \\
Canada}
\maketitle

\begin{abstract}
A time fractional quantum framework has been introduced into quantum
mechanics. A new version of the space-time fractional Schr\"{o}dinger
equation has been launched. The introduced space-time fractional Schr\"{o}%
dinger equation has a new scale parameter, which is a fractional
generalization of Planck's constant in quantum physics.

It has been shown that the presence of a time fractional time derivative in
the space-time fractional Schr\"{o}dinger equation significantly impacts
quantum mechanical fundamentals.

Time fractional quantum mechanical operators of coordinate, momentum and
angular momentum were defined and their commutation relationships were
established. The pseudo-Hamilton operator was introduced and its Hermicity
has been proven.

Two new functions related to the Mittag-Leffler function have been
introduced to solve the space-time fractional Schr\"{o}dinger equation.
Energy of a time fractional quantum system has been defined and calculated
in terms of the newly introduced functions. It has been found that in the
framework of time fractional quantum mechanics there are no stationary
states, and the eigenvalues of the pseudo-Hamilton operator are not the
energy levels of the time fractional quantum system.

A free particle solution to the space-time fractional Schr\"{o}dinger
equation was found. A free particle space-time fractional quantum mechanical
kernel has been found and expressed in terms of the $H$-function.
Renormalization properties of a free particle solution and the space-time
fractional quantum kernel were established.

Some particular cases of time fractional quantum mechanics have been
analyzed and discussed.

\textit{PACS }number(s): 05.40.Fb, 05.30.-d, 03.65.Sq
\end{abstract}

\tableofcontents

\section{Introduction}

In recent years the application of fractional calculus in quantum theory
became a rapidly growing area. It was initiated by the discovery of
fractional quantum mechanics \cite{Laskin1}-\cite{Laskin4}. The crucial
manifestation of fractional quantum mechanics is \textit{fractional Schr\"{o}%
dinger equation}. The fractional Schr\"{o}dinger equation includes a spatial
derivative of fractional order instead of the second order spatial
derivative in the well-known Schr\"{o}dinger equation. Thus, only the
spatial derivative becomes fractional in the fractional Schr\"{o}dinger
equation, while the time derivative is the first order time derivative. Due
to the presence of the first order time derivative in the fractional Schr%
\"{o}dinger equation, fractional quantum mechanics supports all quantum
mechanics fundamentals.

Inspired by the work of Laskin \cite{Laskin1}-\cite{Laskin2}, Naber invented 
\textit{time fractional Schr\"{o}dinger equation }\cite{Naber}. The time
fractional Schr\"{o}dinger equation involves the time derivative of
fractional order instead of the first-order time derivative, while the
spatial derivative is the second-order spatial derivative as it is in the
well-known Schr\"{o}dinger equation. To obtain the time fractional Schr\"{o}%
dinger equation, Naber mapped the time fractional diffusion equation into
the time fractional Schr\"{o}dinger equation, similarly to the map between
the well-known diffusion equation and the standard Schr\"{o}dinger equation.
The mapping implemented by Naber can be considered as a "fractional"
generalization of the Wick rotation \cite{Wick}. To get the time fractional
Schr\"{o}dinger equation, Naber implemented the Wick rotation in complex $t$%
-plane by rising the imaginary unit $i$ to the same fractional power as the
fractional order of the time derivative in the time fractional diffusion
equation. The time fractional derivative in the time fractional Schr\"{o}%
dinger equation is the Caputo fractional derivative \cite{Caputo}. Naber has
found the exact solutions to the time fractional Schr\"{o}dinger equation
for a free particle and a particle in a potential well \cite{Naber}.

Later on, Wang and Xu \cite{WangandXu}, and then Dong and Xu \cite{DongandXu}%
, combined both Laskin's equation and Naber's equation and came up with 
\textit{space-time fractional Schr\"{o}dinger equation}. The space-time
fractional Schr\"{o}dinger equation includes both spatial and temporal
fractional derivatives. Wang and Xu found exact solutions to the space-time
fractional Schr\"{o}dinger equation for a free particle and for an infinite
square potential well. Dong and Xu found the exact solution to the
space-time fractional Schr\"{o}dinger equation for a quantum particle in $%
\delta $-potential well.

Here we introduce time fractional quantum mechanics and develop its
fundamentals. The wording "time fractional quantum mechanics" means that the
time derivative in the fundamental quantum mechanical equations - Schr\"{o}%
dinger equation and fractional Schr\"{o}dinger equation, is substituted with
a fractional time derivative. The time fractional derivative in our approach
is the Caputo fractional derivative.

To introduce and develop time fractional quantum mechanics we begin with our
own version of the space-time fractional Schr\"{o}dinger equation. Our
space-time fractional Schr\"{o}dinger equation involves two scale
dimensional parameters, one of which can be considered as a time fractional
generalization of the famous Planck's constant, while the other one can be
interpreted as a time fractional generalization of the scale parameter
emerging in fractional quantum mechanics \cite{Laskin1}-\cite{Laskin4}. The
fractional generalization of Planck's constant is a fundamental dimensional
parameter of time fractional quantum mechanics, while the fractional
generalization of Laskin's scale parameter \cite{Laskin1}-\cite{Laskin4}
plays a fundamental role in both time fractional quantum mechanics and time
fractional classical mechanics.

In addition to the above mentioned dimensional parameters, time fractional
quantum \ mechanics involves two dimensionless fractality parameters $\alpha 
$, $1<\alpha \leq 2$ and $\beta $, $0<\beta \leq 1$. Parameter $\alpha $ is
the order of the spatial fractional quantum Riesz derivative \cite{Laskin1}
and $\beta $ is the order of the time fractional derivative. In other words, 
$\alpha $ is responsible for modelling \textit{spatial fractality}, while
parameter $\beta $, which is the order of Caputo fractional derivative, is
responsible for modeling \textit{temporal fractality}.

Time fractional quantum mechanical operators of coordinate, momentum and
angular momentum have been introduced and their commutation relationship has
been established. The pseudo-Hamilton quantum mechanical operator has been
introduced and its Hermiticity has been proven in the framework of time
fractional quantum mechanics. The general solution to the space-time
fractional Schr\"{o}dinger equation was found in the case when the
pseudo-Hamilton operator does not depend on time. Energy of a quantum system
in the framework of time fractional quantum mechanics was defined and
calculated in terms of the Mittag-Leffler function. Two new functions
associated with the Mittag-Leffler function have been launched and
elaborated. These two new functions can be considered as a natural
fractional generalization of the well-known trigonometric functions sine and
cosine. A fractional generalization of the celebrated Euler equation was
discovered. A free particle time fractional quantum kernel was calculated in
terms of Fox's $H$-functions.

In the framework of time fractional quantum mechanics at particular choices
of fractality parameters $\alpha $ and $\beta $, we rediscovered the
following fundamental quantum equations:

1. Schr\"{o}dinger equation (Schr\"{o}dinger equation \cite{Schrodinger}), $%
\alpha =2$\ and $\beta =1$;

2. Fractional Schr\"{o}dinger equation (Laskin equation \cite{Laskin4}), $%
1<\alpha \leq 2$\ and $\beta =1$;

3. Time fractional Schr\"{o}dinger equation (Naber equation \cite{Naber}), \ 
$\alpha =2$ and $0<\beta \leq 1$;

4. Space-time fractional Schr\"{o}dinger equation\ (Wang and Xu \cite%
{WangandXu} and Dong and Xu \cite{DongandXu} equation), $1<\alpha \leq 2$\
and $0<\beta \leq 1$.

\subsection{Shortcomings of time fractional quantum mechanics}

While fractional quantum mechanics \cite{Laskin1}-\cite{Laskin4} supports
all quantum mechanics fundamentals, time fractional quantum mechanics
violates the following fundamental physical laws of Quantum Mechanics:

a. Quantum superposition law;

b. Unitarity of evolution operator;

c. Probability conservation law;

d. Existence of stationary energy levels of quantum system.

Fractional quantum dynamics is governed by a pseudo-Hamilton operator
instead of the Hamilton operator in standard quantum mechanics. Eigenvalues
of quantum pseudo-Hamilton operator are not the energy levels of a time
fractional quantum system.

\subsection{Benefits of time fractional quantum mechanics}

What benefits does time fractional quantum mechanics bring into quantum
theory and its applications?

Despite of the above listed shortcomings, the developments in time
fractional quantum mechanics can be considered a newly emerged and
attractive application of fractional calculus to quantum theory. Time
fractional quantum mechanics helps to understand the significance and
importance of the fundamentals of quantum mechanics such as Hamilton
operator, unitarity of evolution operator, existence of stationary energy
levels of quantum mechanical system, quantum superposition law, conservation
of quantum probability, etc.

Besides, time fractional quantum mechanics invokes new mathematical tools,
which have never been used in quantum theory before.

From a stand point of quantum mechanical fundamentals, time fractional
quantum mechanics seems not as a quantum physical theory but rather as an
adequate, convenient mathematical framework, well adjusted to study
dissipative quantum systems interacting with environment \cite{Tarasov1}.

\subsection{The paper outline}

The introduction presents a brief overview of benefits and shortcomings of
time fractional quantum mechanics.

In Sec.2 we launched a new version of the space-time fractional Schr\"{o}%
dinger equation. Our space-time fractional Schr\"{o}dinger equation involves
two scale dimensional parameters, one of which can be considered as a time
fractional generalization of the famous Planck's constant, while the other
one can be interpreted as a time fractional generalization of the scale
parameter emerging in fractional quantum mechanics. A 3D generalization of
the space-time fractional Schr\"{o}dinger equation has been developed. We
also found the space-time fractional Schr\"{o}dinger equation in momentum
representation. The pseudo-Hamilton operator was introduced and its
Hermiticity has been proven. The parity conservation law has been proven in
the framework of time fractional quantum mechanics.

In Sec.3 the solution to the space-time fractional Schr\"{o}dinger equation
was obtained in the case when the pseudo-Hamilton operator does not depend
on time. It was found that time fractional quantum mechanics does not
support a fundamental property of quantum mechanics - conservation of
quantum mechanical probability.

The concept of energy of a time fractional quantum mechanical system was
introduced in Sec.4. It has been shown that in time fractional quantum
mechanics a quantum system does not have stationary states, and eigenvalues
of the pseudo-Hamilton operator are not energy levels.

A free particle wave function and a space-time fractional quantum mechanical
kernel were obtained as applications of time fractional quantum mechanics in
Sec.5. The renormalization properties of a space-time fractional quantum
mechanical \ kernel were established. \ The involvement of the Wright
function in time fractional quantum mechanical framework was discussed.
Fox's $H$-function representation for a free particle space-time fractional
quantum kernel was found.

Sec.6 deals with special cases of time fractional quantum mechanics. It has
been shown that with a particular choice of fractality parameters $\alpha $
and $\beta $, the space-time fractional Schr\"{o}dinger equation covers the
following three special cases: the Schr\"{o}dinger equation ($\alpha =2$\
and $\beta =1$), the fractional Schr\"{o}dinger equation ($1<\alpha \leq 2$\
and $\beta =1$) and the time fractional Schr\"{o}dinger equation ($\alpha =2$
and $0<\beta \leq 1$).

The conclusion summarizes the major findings of the paper.

In Appendix A we introduce and discuss two new functions closely related to
the Mittag-Leffler function. It has been observed that these two new
functions can be considered as a natural fractional generalization of the
well-known trigonometric functions $\cos (z)$, and $\sin (z).$ Fractional
generalization of the celebrated Euler equation was discovered in terms of
these two newly introduced functions.

Appendix B presents the definition and some fundamental properties of Fox's $%
H$-function.

\section{Space-time fractional Schr\"{o}dinger equation}

To introduce time fractional quantum mechanics we begin with our own version
of the space-time fractional Schr\"{o}dinger equation originally introduced
by Wang and Xu \cite{WangandXu} and by Dong and Xu \cite{DongandXu}. The
term \textit{space-time fractional Schr\"{o}dinger equation} was coined by
Dong and Xu \cite{DongandXu}.

Considering a 1D spatial dimension, we launch the space-time fractional Schr%
\"{o}dinger equation in the following form,

\begin{equation}
i^{\beta }\hbar _{\beta }\partial _{t}^{\beta }\psi (x,t)=D_{\alpha ,\beta
}(-\hbar _{\beta }^{2}\Delta )^{\alpha /2}\psi (x,t)+V(x,t)\psi (x,t),
\label{eq1}
\end{equation}

\begin{equation*}
1<\alpha \leq 2,\qquad 0<\beta \leq 1,
\end{equation*}

here $\psi (x,t)$ is the wave function, $i$\ is imanaginery unit, $i=\sqrt{-1%
}$, $V(x,t)$ is potential energy, $\hbar _{\beta }$ and $D_{\alpha ,\beta }$
are two scale coefficients, which we introduce into the framework of time
fractional quantum mechanics, $\Delta $ is the 1D Laplace operator, $\Delta
=\partial ^{2}/\partial x^{2}$, and, finally, $\partial _{t}^{\beta }$ is
the left Caputo fractional derivative \cite{Caputo} of order $\beta $
defined by

\begin{equation}
\partial _{t}^{\beta }f(t)=\frac{1}{\Gamma (1-\beta )}\int\limits_{0}^{t}d%
\tau \frac{f^{^{\prime }}(\tau )}{(t-\tau )^{\beta }},\qquad 0<\beta \leq 1,
\label{eq2}
\end{equation}

where $f^{^{\prime }}(\tau )$ is first order time derivative, $\ f^{^{\prime
}}(\tau )=df(\tau )/d\tau $, $\Gamma (1-\beta )$ is the Gamma function%
\footnote{%
The Gamma function has familiar representaion
\par
\begin{equation}
\Gamma (z)=\int\limits_{0}^{\infty }dte^{-t}t^{z-1}.  \label{eq3}
\end{equation}%
}, and the operator $(-\hbar _{\beta }^{2}\Delta )^{\alpha /2}$ is a time
fractional quantum Riesz derivative\footnote{%
Quantum Riesz fractional derivative was originally introduced in \cite%
{Laskin1}.}

\begin{equation}
(-\hbar _{\beta }^{2}\Delta )^{\alpha /2}\psi (x,t)=\frac{1}{2\pi \hbar
_{\beta }}\int dp_{\beta }\exp \{i\frac{p_{\beta }x}{\hbar _{\beta }}%
\}|p_{\beta }|^{\alpha }\varphi (p_{\beta },t),  \label{eq4}
\end{equation}

where the wave functions in space and momentum representations, $\psi (x,t)$
and $\varphi (p,t)$, are related to each other by the Fourier transforms

\begin{equation}
\psi (x,t)=\frac{1}{2\pi \hbar _{\beta }}\int dp_{\beta }\exp \{i\frac{%
p_{\beta }x}{\hbar _{\beta }}\}\varphi (p_{\beta },t),  \label{eq5}
\end{equation}

\begin{equation}
\varphi (p_{\beta },t)=\int dr\exp \{-i\frac{p_{\beta }x}{\hbar _{\beta }}%
\}\psi (x,t).  \label{eq6}
\end{equation}

Quantum scale coefficient $\hbar _{\beta }$ has physical dimension%
\begin{equation}
\lbrack \hbar _{\beta }]=\mathrm{erg}\cdot \sec ^{\beta },  \label{eq7}
\end{equation}

and scale coefficient $D_{\alpha ,\beta }$ has physical dimension

\begin{equation}
\lbrack D_{\alpha ,\beta }]=\mathrm{erg}^{1-\alpha }\cdot \mathrm{cm}%
^{\alpha }\cdot \mathrm{sec}^{-\alpha \beta }.  \label{eq8}
\end{equation}

The introduction of the scale coefficient $D_{\alpha ,\beta }$ was inspired
by Bayin \cite{Selcuk}.

\subsection{Pseudo-Hamilton operator in time fractional quantum mechanics}

Aiming to obtain the operator form of space-time fractional Schr\"{o}dinger
equation Eq.(\ref{eq1}), let us define time fractional quantum momentum
operator $\widehat{p}_{\beta }$

\begin{equation}
\widehat{p}_{\beta }=-i\hbar _{\beta }\frac{\partial }{\partial x},
\label{eq9}
\end{equation}

and quantum operator of coordinate $\widehat{x}$,

\begin{equation}
\widehat{x}=x.  \label{eq10}
\end{equation}

Hence, the commutation relation in the framework of time fractional quantum
mechanics has the form

\begin{equation}
\lbrack \widehat{x},\widehat{p}_{\beta }]=i\hbar _{\beta },  \label{eq11}
\end{equation}

where $[\widehat{x},\widehat{p}_{\beta }]$ is the commutator of two quantum
operators $\widehat{x}$ and $\widehat{p}_{\beta }$,

\begin{equation}
\lbrack \widehat{x},\widehat{p}_{\beta }]=\widehat{x}\widehat{p}_{\beta }-%
\widehat{p}_{\beta }\widehat{x}.  \label{eq12}
\end{equation}

Using quantum operators of momentum Eq.(\ref{eq9}) and coordinate Eq.(\ref%
{eq10}), we introduce a new time fractional quantum mechanical operator $%
\widehat{H}_{\alpha ,\beta }(\widehat{p}_{\beta },\widehat{x})$, 
\begin{equation}
\widehat{H}_{\alpha ,\beta }(\widehat{p}_{\beta },\widehat{x})=D_{\alpha
,\beta }|\widehat{p}_{\beta }|^{\alpha }+V(\widehat{x},t).  \label{eq13}
\end{equation}

The operator $\widehat{H}_{\alpha ,\beta }(\widehat{p}_{\beta },\widehat{x})$
is not the Hamilton operator of the quantum mechanical system under
consideration\footnote{%
In time fractional quantum mechanics the eigenvalues of operator $\widehat{H}%
_{\alpha ,\beta }(\widehat{p}_{\beta },\widehat{x})$ are not energies of a
quantum system. In classical time fractional mechanics the principle of
least action with classical pseudo-Hamilton function $H_{\alpha ,\beta
}(p_{\beta },x)$ does not result in Hamilton equations of motion.}.
Following \cite{DongandXu}, we will call this operator \textit{%
pseudo-Hamilton operator}.

Having the pseudo-Hamilton operator $\widehat{H}_{\alpha ,\beta }$, let us
rewrite Eq.(\ref{eq1}) as

\begin{equation}
\hbar _{\beta }i^{\beta }\partial _{t}^{\beta }\psi (x,t)=\widehat{H}%
_{\alpha ,\beta }(\widehat{p}_{\beta },\widehat{x})\psi (x,t)  \label{eq14}
\end{equation}

\begin{equation*}
=\left( D_{\alpha ,\beta }|\widehat{p}_{\beta }|^{\alpha }+V(\widehat{x}%
,t)\right) \psi (x,t),\qquad 1<\alpha \leq 2,\qquad 0<\beta \leq 1,
\end{equation*}

which is operator form of the space-time fractional Schr\"{o}dinger equation.

\subsection{Hermiticity of pseudo-Hamilton operator}

The pseudo-Hamilton operator $\widehat{H}_{\alpha ,\beta }(\widehat{p}%
_{\beta },\widehat{x})$ introduced by Eq.(\ref{eq13}) is the Hermitian
operator in space with scalar product

\begin{equation}
(\phi ,\chi )=\int\limits_{-\infty }^{\infty }dx\phi ^{\ast }(x,t)\chi (x,t).
\label{eq15}
\end{equation}

In this space, operators $\widehat{p}_{\beta }$ and $\widehat{x}$ defined by
Eqs.(\ref{eq9}) and (\ref{eq10}) are Hermitian operators. The proof can be
found in any textbook on quantum mechanics, (see, for example, \cite{Landau}%
).

To prove the Hermiticity of quantum mechanical operator $\widehat{H}_{\alpha
,\beta }$ let us note that in accordance with the definition of the time
fractional quantum Riesz derivative given by Eq.(\ref{eq4}) there exists the
integration-by parts formula

\begin{equation}
(\phi ,(-\hbar _{\beta }^{2}\Delta )^{\alpha /2}\chi )=((-\hbar _{\beta
}^{2}\Delta )^{\alpha /2}\phi ,\chi ).  \label{eq16}
\end{equation}

Therefore, using this integration-by parts formula we prove
straightforwardly Hermiticity of the term $D_{\alpha ,\beta }|\widehat{p}%
_{\beta }|^{\alpha }=D_{\alpha ,\beta }(-\hbar _{\beta }^{2}\Delta )^{\alpha
/2}$. Next, potential energy operator $V(\widehat{x},t)$ in Eq.(\ref{eq13})
is Hermitian operator by virtue of being a function of Hermitian operator $%
\widehat{x}$.

Thus, we completed the proof of Hermiticity of the pseudo-Hamilton operator $%
\widehat{H}_{\alpha ,\beta }$ in the space with scalar product defined by
Eq.(\ref{eq15}).

\subsection{3D generalization of space-time fractional Schr\"{o}dinger
equation}

Considering 3D spatial dimension, we launch the space-time fractional Schr%
\"{o}dinger equation of the following form,

\begin{equation}
i^{\beta }\hbar _{\beta }\partial _{t}^{\beta }\psi (\mathbf{r},t)=D_{\alpha
,\beta }(-\hbar _{\beta }^{2}\Delta )^{\alpha /2}\psi (\mathbf{r},t)+V(%
\mathbf{r},t)\psi (\mathbf{r},t),  \label{eq17}
\end{equation}

\begin{equation*}
1<\alpha \leq 2,\qquad 0<\beta \leq 1,
\end{equation*}

where $\psi (\mathbf{r},t)$ is the wave function, $\mathbf{r}$\ is 3D space
vector, $\Delta $ is the Laplacian, $\Delta =(\partial /\partial \mathbf{r)}%
^{2},$ all other notations are the same as for Eq.(\ref{eq1}), and 3D time
fractional quantum Riesz derivative $(-\hbar _{\beta }^{2}\Delta )^{\alpha
/2}$ is defined by

\begin{equation}
(-\hbar _{\beta }^{2}\Delta )^{\alpha /2}\psi (\mathbf{r},t)=\frac{1}{(2\pi
\hbar _{\beta })^{3}}\int d^{3}r\exp \{i\frac{\mathbf{p}_{\beta }\mathbf{r}}{%
\hbar _{\beta }}\}|\mathbf{p}_{\beta }|^{\alpha }\varphi (\mathbf{p}_{\beta
},t),  \label{eq18}
\end{equation}

where the wave functions in space representation $\psi (\mathbf{r},t)$ and
momentum representation $\varphi (\mathbf{p},t)$ are related to each other
by the 3D Fourier transforms

\begin{equation}
\psi (\mathbf{r},t)=\frac{1}{(2\pi \hbar _{\beta })^{3}}\int d^{3}p\exp \{i%
\frac{\mathbf{p}_{\beta }\mathbf{r}}{\hbar _{\beta }}\}\varphi (\mathbf{p}%
_{\beta },t),  \label{eq19}
\end{equation}

\begin{equation}
\varphi (\mathbf{p}_{\beta },t)=\int d^{3}r\exp \{-i\frac{\mathbf{p}_{\beta }%
\mathbf{r}}{\hbar _{\beta }}\}\psi (\mathbf{r},t).  \label{eq20}
\end{equation}

Further, 3D generalization of the pseudo-Hamilton operator $\widehat{H}%
_{\alpha ,\beta }$ in the framework of time fractional quantum mechanics is

\begin{equation}
\widehat{H}_{\alpha ,\beta }(\widehat{\mathbf{p}}_{\beta },\widehat{\mathbf{r%
}})=D_{\alpha ,\beta }|\widehat{\mathbf{p}}_{\beta }|^{\alpha }+V(\widehat{%
\mathbf{r}},t),  \label{eq21}
\end{equation}

where $\widehat{\mathbf{r}}$ is 3D quantum operator of coordinate

\begin{equation}
\widehat{\mathbf{r}}=\mathbf{r,}  \label{eq22}
\end{equation}

and $\widehat{\mathbf{p}}_{\beta }$ is 3D time fractional quantum momentum
operator introduced by

\begin{equation}
\widehat{\mathbf{p}}_{\beta }=-i\hbar _{\beta }\frac{\partial }{\partial 
\mathbf{r}},  \label{eq23}
\end{equation}

with $\hbar _{\beta }$ being the scale coefficient appearing for the first
time in Eqs.(\ref{eq1}) and (\ref{eq17}).

The basic canonical commutation relationships in 3D case are

\begin{equation}
\lbrack \widehat{r}_{k},\widehat{p}_{\beta j}]=i\hbar _{\beta }\delta
_{kj},\qquad \lbrack \widehat{r}_{k},\widehat{r}_{j}]=0,\qquad \lbrack 
\widehat{p}_{\beta k},\widehat{p}_{\beta j}]=0,\qquad k,j=1,2,3,
\label{eq24}
\end{equation}

where $\delta _{kj}$ is the Kronecker symbol,

\begin{equation}
\delta _{kj},={\LARGE \{}\QATOP{1\qquad k=j,}{0\qquad k\neq j.}  \label{eq25}
\end{equation}

Having 3D generalization of pseudo-Hamilton operator $\widehat{H}_{\alpha
,\beta }(\widehat{\mathbf{p}}_{\beta },\widehat{\mathbf{r}})$, let us
present Eq.(\ref{eq17}) in the form

\begin{equation}
\hbar _{\beta }i^{\beta }\partial _{t}^{\beta }\psi (\mathbf{r},t)=\widehat{H%
}_{\alpha ,\beta }(\widehat{\mathbf{p}}_{\beta },\widehat{\mathbf{r}})\psi (%
\mathbf{r},t)  \label{eq26}
\end{equation}

\begin{equation*}
=\left( D_{\alpha ,\beta }|\widehat{\mathbf{p}}_{\beta }|^{\alpha }+V(%
\widehat{\mathbf{r}},t)\right) \psi (\mathbf{r},t),\qquad 1<\alpha \leq
2,\qquad 0<\beta \leq 1,
\end{equation*}

which is operator form of 3D space-time fractional Schr\"{o}dinger equation.

Heaving operators (\ref{eq22}) and (\ref{eq23}) we introduce time fractional
angular momentum operator $\widehat{\mathbf{L}}_{\beta }$ as cross-product
of two above defined operators $\widehat{\mathbf{r}}$ and $\widehat{\mathbf{p%
}}_{\beta }$

\begin{equation}
\widehat{\mathbf{L}}_{\beta }=\widehat{\mathbf{r}}\times \widehat{\mathbf{p}}%
_{\beta }=-i\hbar _{\beta }\mathbf{r}\times \frac{\partial }{\partial 
\mathbf{r}}.  \label{eq27}
\end{equation}

This equation can be expressed in component form

\begin{equation}
\widehat{\mathrm{L}}_{\beta i}=\varepsilon _{ijk}\widehat{r}_{j}\widehat{p}%
_{\beta k},\qquad i,k,j=1,2,3,  \label{eq28}
\end{equation}

if we use 3D Levi-Civita antisymmetric tensor $\varepsilon _{ijk\text{ }}$
which changes its sign under interchange of any pair of indices $i$, $j$, $k$%
.

It is obvious that the algebra of time fractional angular momentum operators 
$\widehat{\mathrm{L}}_{\beta i}$ and its commutation relationships with
operators $\widehat{r}_{j}$ and $\widehat{p}_{\beta k}$ are the same as for
the angular momentum operator of quantum mechanics \cite{Landau}.

\subsubsection{Space-time fractional Schr\"{o}dinger equation in momentum
representation}

To obtain the space-time fractional Schr\"{o}dinger equation in momentum
representation let us substitute the wave function $\psi (\mathbf{r},t)$
from Eq.(\ref{eq19}) into Eq.(\ref{eq17}),

\begin{equation}
i^{\beta }\hbar _{\beta }\partial _{t}^{\beta }\frac{1}{(2\pi \hbar _{\beta
})^{3}}\int d^{3}p_{\beta }^{\prime }\exp \{i\frac{\mathbf{p}_{\beta
}^{\prime }\mathbf{r}}{\hbar _{\beta }}\}\varphi (\mathbf{p}_{\beta
}^{\prime },t)  \label{eq29}
\end{equation}

\begin{equation*}
=\frac{D_{\alpha ,\beta }}{(2\pi \hbar _{\beta })^{3}}\int d^{3}p_{\beta
}^{\prime }\exp \{i\frac{\mathbf{p}_{\beta }^{\prime }\mathbf{r}}{\hbar
_{\beta }}\}|\mathbf{p}_{\beta }^{\prime }|^{\alpha }\varphi (\mathbf{p}%
_{\beta }^{\prime },t)+\frac{V(\mathbf{r,}t)}{(2\pi \hbar )^{3}}\int
d^{3}p_{\beta }^{\prime }\exp \{i\frac{\mathbf{p}_{\beta }^{\prime }\mathbf{r%
}}{\hbar _{\beta }}\}\varphi (\mathbf{p}_{\beta }^{\prime },t).
\end{equation*}

Further, multiplying Eq.(\ref{eq29}) by $\exp (-i\mathbf{p}_{\beta }\mathbf{r%
}/\hbar _{\beta })$ and integrating over $d^{3}r$ yields the equation for
the wave function $\varphi (\mathbf{p}_{\beta },t)$ in momentum
representation

\begin{equation}
i^{\beta }\hbar _{\beta }\partial _{t}^{\beta }\varphi (\mathbf{p}_{\beta
},t)=D_{\alpha ,\beta }|\mathbf{p}_{\beta }|^{\alpha }\varphi (\mathbf{p}%
_{\beta },t)+\int d^{3}p_{\beta }^{\prime }U_{\mathbf{p}_{\beta },\mathbf{p}%
_{\beta }^{\prime }}\varphi (\mathbf{p}_{\beta }^{\prime },t),  \label{eq30}
\end{equation}

\begin{equation*}
1<\alpha \leq 2,\qquad 0<\beta \leq 1,
\end{equation*}

where $U_{\mathbf{p}_{\beta },\mathbf{p}_{\beta }^{\prime }}$ is introduced
as

\begin{equation}
U_{\mathbf{p}_{\beta },\mathbf{p}_{\beta }^{\prime }}=\frac{1}{(2\pi \hbar
_{\beta })^{3}}\int d^{3}r\exp (-i(\mathbf{p}_{\beta }\mathbf{-p}_{\beta
}^{\prime })\mathbf{r}/\hbar _{\beta })V(\mathbf{r,}t),  \label{eq31}
\end{equation}

and we used the following representation for the delta function $\delta (%
\mathbf{r})$

\begin{equation}
\delta (\mathbf{r})=\frac{1}{(2\pi \hbar _{\beta })^{3}}\int d^{3}p_{\beta
}\exp (i\mathbf{p}_{\beta }\mathbf{r}/\hbar _{\beta }).  \label{eq32}
\end{equation}

Equation (\ref{eq30}) is the 3D space-time fractional Schr\"{o}dinger
equation in momentum representation.

Substituting the wave function $\psi (x,t)$ from Eq.(\ref{eq5}) into Eq.(\ref%
{eq1}), multiplying by $\exp (-ipx/\hbar )$ and integrating over $dx$ bring
us the 1D space-time fractional Schr\"{o}dinger equation in momentum
representation

\begin{equation}
i^{\beta }\hbar _{\beta }\partial _{t}^{\beta }\varphi (p_{\beta
},t)=D_{\alpha ,\beta }|p_{\beta }|^{\alpha }\varphi (p_{\beta },t)+\int
d^{3}p_{\beta }^{\prime }U_{p_{\beta },p_{\beta }^{\prime }}\varphi
(p_{\beta }^{\prime },t),  \label{eq33}
\end{equation}

\begin{equation*}
1<\alpha \leq 2,\qquad 0<\beta \leq 1,
\end{equation*}

where $U_{p_{\beta },p_{\beta }^{\prime }}$ is given by

\begin{equation}
U_{p_{\beta },p_{\beta }^{\prime }}=\frac{1}{2\pi \hbar _{\beta }}\int
dx\exp (-i(p_{\beta }\mathbf{-}p_{\beta }^{\prime })x/\hbar _{\beta })V(x),
\label{eq34}
\end{equation}

and we used the following representation for the delta function $\delta (x)$

\begin{equation}
\delta (x)=\frac{1}{2\pi \hbar _{\beta }}\int dp_{\beta }\exp (ip_{\beta
}x/\hbar _{\beta }).  \label{eq35}
\end{equation}

Equation (\ref{eq30}) is the 1D space-time fractional Schr\"{o}dinger
equation in momentum representation for the wave function $\varphi (p_{\beta
},t)$.

\subsubsection{Hermiticity of pseudo-Hamilton operator $\widehat{H}_{\protect%
\alpha ,\protect\beta }(\widehat{\mathbf{p}}_{\protect\beta },\widehat{%
\mathbf{r}})$}

To prove that defined by Eq.(\ref{eq17}) $\widehat{H}_{\alpha ,\beta }(%
\widehat{\mathbf{p}}_{\beta },\widehat{\mathbf{r}})$ is Hermitian operator
we introduce the space with scalar product defined by

\begin{equation}
(\phi (\mathbf{r},t),\chi (\mathbf{r},t))=\int d\mathbf{r}\phi ^{\ast }(%
\mathbf{r},t)\chi (\mathbf{r},t),  \label{eq36}
\end{equation}

where $\phi ^{\ast }(\mathbf{r},t)$ means complex conjugate function.

Then, the proof can be done straightforwardly by 3D generalization of the
considered presented in Sec.2.2.

\subsubsection{The parity conservation law}

Here we study invariance of pseudo-Hamilton operator $\widehat{H}_{\alpha
,\beta }(\widehat{\mathbf{p}}_{\beta },\widehat{\mathbf{r}})$ under \textit{%
inversion} transformation. Inversion, or to be precise, spatial inversion
consists of the simultaneous change in the sign of all three spatial
coordinates

\begin{equation}
\mathbf{r}\rightarrow -\mathbf{r},\qquad x\rightarrow -x,\quad y\rightarrow
-y,\quad z\rightarrow -z.  \label{eq37}
\end{equation}

It easy to see that

\begin{equation}
(-\hbar _{\beta }^{2}\Delta )^{\alpha /2}\exp \{i\frac{\mathbf{p}_{\beta }%
\mathbf{x}}{\hbar _{\beta }}\}=|\mathbf{p}_{\beta }|^{\alpha }\exp \{i\frac{%
\mathbf{p}_{\beta }\mathbf{x}}{\hbar _{\beta }}\},  \label{eq38}
\end{equation}

which means that the function $\exp \{i\mathbf{p}_{\beta }\mathbf{x}/\hbar
_{\beta }\}$ is the eigenfunction of the 3D time fractional quantum Riesz
operator $(-\hbar _{\beta }^{2}\Delta )^{\alpha /2}$ with eigenvalue $|%
\mathbf{p}_{\beta }|^{\alpha }$.

Thus, the operator $(-\hbar _{\beta }^{2}\Delta )^{\alpha /2}$ is the
symmetrized fractional derivative, that is

\begin{equation}
(-\hbar _{\beta }^{2}\Delta _{\mathbf{r}})^{\alpha /2}...=(-\hbar _{\beta
}^{2}\Delta _{-\mathbf{r}})^{\alpha /2}....  \label{eq39}
\end{equation}

Assuming that the potential energy operator $V(\widehat{\mathbf{r}},t)$ is
invariant under spatial inversion $V(\widehat{\mathbf{r}},t)=V(-\widehat{%
\mathbf{r}},t)$, we conclude that pseudo-Hamilton operator $\widehat{H}%
_{\alpha ,\beta }(\widehat{\mathbf{p}}_{\beta },\widehat{\mathbf{r}})$ is
invariant under inversion, or, in other words, it supports the parity
conservation law. The inverse symmetry results in the fact that inversion
operator $\widehat{P}_{\mathrm{inv}}$ and the pseudo-Hamilton operator $%
\widehat{H}_{\alpha ,\beta }(\widehat{\mathbf{p}}_{\beta },\widehat{\mathbf{r%
}})$ commute

\begin{equation}
\widehat{P}_{\mathrm{inv}}\widehat{H}_{\alpha ,\beta }=\widehat{H}_{\alpha
,\beta }\widehat{P}_{\mathrm{inv}}.  \label{eq40}
\end{equation}

Hence, we can divide the wave functions of time fractional quantum
mechanical states with defined eigenvalue of the operator $\widehat{P}_{%
\mathrm{inv}}$ into two classes; (i) wave functions which are not changed
upon the action of the inversion operator, $\widehat{P}_{\mathrm{inv}}\psi
_{+}(\mathbf{r})=\psi _{+}(\mathbf{r})$, the corresponding states are called
even states; (ii) wave functions which change sign under action of the
inversion operator, $\widehat{P}_{\mathrm{inv}}\psi _{-}(\mathbf{r})=-\psi
_{-}(\mathbf{r})$, the corresponding states are called odd states. Equation(%
\ref{eq40}) represents the parity conservation law\ for time fractional
quantum mechanics, that is, if the state of a time fractional quantum
mechanical system has a given parity (i.e. if it is even, or odd), then this
parity is conserved.

Thus, we conclude that time fractional quantum mechanics supports the parity
conservation law.

\section{Solution to space-time fractional Schr\"{o}dinger equation}

It is well know that if the Hamilton operator of a quantum mechanical system
does not depend on time, then we can search for the solution to the Schr\"{o}%
dinger equation in separable form. In the case when pseudo-Hamilton operator
(\ref{eq21}) doesn't depend on time we search for the solution to Eq.(\ref%
{eq1}) assuming that the solution has form,

\begin{equation}
\psi (x,t)=\varphi (x)\chi (t),  \label{eq41}
\end{equation}

where $\varphi (x)$ and $\chi (t)$ are spatial and temporal components of
the wave function $\psi (x,t)$\footnote{%
We consider here 1D space-time fractional Schr\"{o}dinger equation The
generalization to 3D case can be done straightforwardly.}.

It is assumed as well, that initial wave function $\psi (x,t=0)=\psi (x,0)$
is normalized

\begin{equation}
\int\limits_{-\infty }^{\infty }dx|\psi (x,0)|^{2}=|\chi
(0)|^{2}\int\limits_{-\infty }^{\infty }dx|\varphi (x)|^{2}=1.  \label{eq42}
\end{equation}

Substituting Eq.(\ref{eq41}) into Eq.(\ref{eq1}) we obtain two equations%
\begin{equation}
\widehat{H}_{\alpha ,\beta }(\widehat{p}_{\beta },\widehat{x})\varphi (x)=%
\mathcal{E}\varphi (x),  \label{eq43}
\end{equation}

and

\begin{equation}
i^{\beta }\hbar _{\beta }\partial _{t}^{\beta }\chi (t)=\mathcal{E}\chi
(t),\qquad \chi (t=0)=\chi (0),  \label{eq44}
\end{equation}

where $\mathcal{E}$ is the eigenvalue of quantum mechanical pseudo-Hamilton
operator $\widehat{H}_{\beta }(\widehat{p}_{\beta },\widehat{x})$,

\begin{equation}
\widehat{H}_{\alpha ,\beta }(\widehat{p}_{\beta },\widehat{x})=D_{\alpha
,\beta }|\widehat{p}_{\beta }|^{\alpha }+V(\widehat{x}),\qquad 1<\alpha \leq
2,\qquad 0<\beta \leq 1.  \label{eq45}
\end{equation}

The solution to Eq.(\ref{eq43}) depending on potential energy term $V(%
\widehat{x})$ can be obtained by the well-known methods of standard quantum
mechanics \cite{Landau}. To find the solution to time fractional equation (%
\ref{eq44}) we can use the Laplace transform method\footnote{%
The Laplace transform $\widetilde{\chi }(s)$ of a function $\chi (t)$ is
defined as
\par
\begin{equation}
\widetilde{\chi }(s)=\int\limits_{0}^{\infty }dte^{-st}\chi (t).
\label{eq46}
\end{equation}%
\par
where $\chi (t)$ is defined for $t\geq 0$ if the integral exists.
\par
The inverse Laplace transform is defined by
\par
\begin{equation}
\chi (t)=\frac{1}{2\pi i}\int\limits_{\gamma i-\infty }^{\gamma +i\infty
}dse^{st}\widetilde{\chi }(s),  \label{eq47}
\end{equation}%
where the integration is done along the vertical line $\func{Re}(s)=\gamma $
in the complex plane $s$ such that $\gamma $ is greater than the real part
of all singularities in the complex plane of $\widetilde{\chi }(s)$. This
requirement on integration path in the complex plane $s$ ensures that the
contour path is in the region of convergence. In practice, computing the
complex integral can be done by using the Cauchy residue theorem.}. In the
Laplace transform domain Eq.(\ref{eq44}) reads

\begin{equation}
i^{\beta }\hbar _{\beta }\left( s^{\beta }\widetilde{\chi }(s)-s^{\beta
-1}\chi (0)\right) =\mathcal{E}\widetilde{\chi }(s),  \label{eq48}
\end{equation}

where $\widetilde{\chi }(s)$ is defined by Eq.(\ref{eq46}) and $\chi
(0)=\chi (t=0)$ is the initial condition on time dependent component $\chi
(t)$ of the wave function $\psi (x,t)$ given by Eq.(\ref{eq41})$.$

Further, from Eq.(\ref{eq48}) we have

\begin{equation}
\widetilde{\chi }(s)=\frac{s^{-1}}{1-\mathcal{E}s^{-\beta }/i^{\beta }\hbar
_{\beta }}\chi (0),  \label{eq49}
\end{equation}

which can be presented as geometric series (the range of convergence is
given by the criteria $|\mathcal{E}s^{-\beta }/i^{\beta }\hbar _{\beta }|<1$%
, see, \cite{Selcuk}),

\begin{equation}
\widetilde{\chi }(s)=\chi (0)\sum\limits_{m=0}^{\infty }(\frac{\mathcal{E}%
s^{-\beta }}{i^{\beta }\hbar _{\beta }})^{m}s^{-1}=\chi
(0)\sum\limits_{m=0}^{\infty }(\frac{\mathcal{E}}{i^{\beta }\hbar _{\beta }}%
)^{m}s^{-m\beta -1}.  \label{eq50}
\end{equation}

Then the inverse Laplace transform yields

\begin{equation}
\chi (t)=\chi (0)\sum\limits_{m=0}^{\infty }(\frac{\mathcal{E}}{i^{\beta
}\hbar _{\beta }})^{m}\frac{t^{\beta m}}{\Gamma (\beta m+1)}=\chi
(0)E_{\beta }(\frac{\mathcal{(-}it^{)\beta }\mathcal{E}}{\hbar _{\beta }}),
\label{eq51}
\end{equation}

where $E_{\beta }(z)$ is the Mittag-Leffler function \cite{Mittag-Leffler}
defined by the series

\begin{equation}
E_{\beta }(z)=\sum\limits_{m=0}^{\infty }\frac{z^{m}}{\Gamma (\beta m+1)},
\label{eq52}
\end{equation}

with $\Gamma (x)$ being the Gamma function given by Eq.(\ref{eq3}).

In the limit case, when $\beta =1$, the Mittag-Leffler function $E_{\beta
}(z)$ becomes the exponential function

\begin{equation}
E_{\beta }(z)|_{\beta =1}=\sum\limits_{m=0}^{\infty }\frac{z^{m}}{m!}=\exp
(z),  \label{eq53}
\end{equation}

and $\chi (t)$ goes into

\begin{equation}
\chi (t)|_{\beta =1}=\chi (0)\exp (-i\frac{\mathcal{E}t}{\hbar }),
\label{eq54}
\end{equation}

here $\hbar $ is Planck's constant and $\mathcal{E}$ is an eigenvalue of
quantum mechanical Hamilton operator $\widehat{H}_{\alpha }(\widehat{p},%
\widehat{x}),$

\begin{equation}
\widehat{H}_{\alpha }(\widehat{p},\widehat{x})=\widehat{H}_{\alpha ,\beta }(%
\widehat{p}_{\beta },\widehat{x})|_{\beta =1}=D_{\alpha }|\widehat{p}%
|^{\alpha }+V(\widehat{x}),\qquad 1<\alpha \leq 2,  \label{eq55}
\end{equation}

where the scale coefficient $D_{\alpha }=D_{\alpha ,\beta }|_{\beta =1}$ was
introduced originally in \cite{Laskin1}.

It is easy to see, that the time dependent component $\chi (t)$ (\ref{eq51})
of the wave function $\psi (x,t)$ can be written as

\begin{equation}
\chi (t)=\chi (0)\left\{ Ec_{\beta }(\frac{\mathcal{E(-}t)^{\beta }}{\hbar
_{\beta }})+i^{\beta }Es_{\beta }(\frac{\mathcal{E(-}t)^{\beta }}{\hbar
_{\beta }})\right\} ,  \label{eq56}
\end{equation}

if we use the following new expression for the Mittag-Leffler function $%
E_{\beta }(i^{\beta }z)$,

\begin{equation}
E_{\beta }(i^{\beta }z)=Ec_{\beta }(z)+i^{\beta }Es_{\beta }(z),\qquad
0<\beta \leq 1,  \label{eq57}
\end{equation}

in \ terms of two functions $Ec_{\beta }(z)$ and $Es_{\beta }(z)$ introduced
by Eqs.(\ref{eq13_sol}) and (\ref{eq14_sol}), see Appendix A.

It has been shown in Appendix A that Eq.(\ref{eq57}) can be considered as a
fractional generalization of the celebrated Euler equation, which is
recovered from Eq.(\ref{eq57}) in the limit case $\beta =1.$

Finally, we have the solution to the time fractional Schr\"{o}dinger
equation (\ref{eq1}) given by

\begin{equation}
\psi (x,t)=\varphi (x)E_{\beta }(\frac{\mathcal{E}t^{\beta }}{i^{\beta
}\hbar _{\beta }})=\varphi (x)\chi (0)\left\{ Ec_{\beta }(\frac{\mathcal{E(-}%
t)^{\beta }}{\hbar _{\beta }})+i^{\beta }Es_{\beta }(\frac{\mathcal{E(-}%
t)^{\beta }}{\hbar _{\beta }})\right\} .  \label{eq58}
\end{equation}

We see from Eq.(\ref{eq58}) that the time fractional quantum mechanics does
not support normalization condition for the wave function. If normalization
condition (\ref{eq42}) holds at the initial time moment $t=0$, then at any
time moment $t>0$\ it becomes time dependent

\begin{equation}
\int\limits_{-\infty }^{\infty }dx|\psi (x,t)|^{2}=|Ec_{\beta }(\frac{%
\mathcal{E(-}t^{\beta })}{\hbar _{\beta }})+i^{\beta }Es_{\beta }(\frac{%
\mathcal{E(-}t^{\beta })}{\hbar _{\beta }})|^{2}\cdot |\chi
(0)|^{2}\int\limits_{-\infty }^{\infty }dx|\varphi (x)|^{2}  \label{eq59}
\end{equation}

\begin{equation*}
=\left( Ec_{\beta }^{2}(\frac{\mathcal{E(-}t^{\beta })}{\hbar _{\beta }}%
)+Es_{\beta }^{2}(\frac{\mathcal{E(-}t^{\beta })}{\hbar _{\beta }})+2\cos 
\frac{\pi \beta }{2}Ec_{\beta }(\frac{\mathcal{E(-}t^{\beta })}{\hbar
_{\beta }})Es_{\beta }(\frac{\mathcal{E(-}t^{\beta })}{\hbar _{\beta }}%
)\right) ,
\end{equation*}

here normalization condition Eq.(\ref{eq42}) was used.

Therefore, we come to the conclusion that in the framework of time
fractional quantum mechanics total quantum mechanical probability $%
\int\limits_{-\infty }^{\infty }dx|\psi (x,t)|^{2}$ is time dependent. In
other words, \ time fractional quantum mechanics does not support a
fundamental property of the quantum mechanics - conservation of quantum
mechanical probability.

\section{Energy in the framework of time fractional quantum mechanics}

It is well known that the energy levels of a quantum mechanical system in
stationary states are defined by Hamilton operator eigenvalues. In time
fractional quantum mechanics a quantum system doesn't have stationary states
and an eigenvalues of pseudo-Hamilton operator are not energy levels.

To define and calculate what we call as \textit{energy of time fractional
quantum system}, let us introduce \textit{time fractional quantum mechanical
operator of energy} follow to Bayin \cite{Selcuk},

\begin{equation}
\widehat{\mathrm{E}}_{\beta }=i^{\beta }\hbar _{\beta }\partial _{t}^{\beta
},\qquad 0<\beta \leq 1.  \label{eq2sh_ener}
\end{equation}

When $\beta =1$, the Caputo fractional derivative Eq.(\ref{eq2}) becomes the
ordinary first order time derivative, the quantum scale coefficient $\hbar
_{\beta }$ becomes the well-known fundamental Planck's constant, $\hbar
_{\beta }|_{\beta =1}=\hbar $. Thus, in the case $\beta =1$ time fractional
quantum mechanical operator of energy (\ref{eq2sh_ener}) becomes what is
called sometimes as operator of energy,

\begin{equation}
\widehat{\mathrm{E}}_{\beta }|_{\beta =1}=\widehat{\mathrm{E}}=i\hbar
\partial _{t}.  \label{eq2sh-ener}
\end{equation}

However, it has to be clear that time is not quantum mechanical operator in
the framework of time fractional quantum mechanics as well as in the
framework of standard quantum mechanics.

Having energy operator $\widehat{\mathrm{E}}_{\beta }$ (\ref{eq2sh_ener})
and wave function given by Eq.(\ref{eq58}) we can calculate the energy of
time fractional quantum mechanical system.

\begin{equation}
\mathrm{E}_{\beta }=\int dx\psi ^{\ast }(x,t)\widehat{\mathrm{E}}_{\beta
}\psi (x,t)=i^{\beta }\hbar _{\beta }\int dx\psi ^{\ast }(x,t)\partial
_{t}^{\beta }\psi (x,t),  \label{eq3sh-ener}
\end{equation}

where $\partial _{t}^{\beta }$ is the Caputo fractional time derivative (\ref%
{eq2}) and $\psi ^{\ast }(x,t)$ stands for complex conjugated wave function.

Substituting $\psi (x,t)$ given by Eq.(\ref{eq41}) with $\chi (t)$ defined
by Eq.(\ref{eq56}) yields for $\mathrm{E}_{\beta }$

\begin{equation}
\mathrm{E}_{\beta }=\mathcal{E}|\chi (0)|^{2}\cdot |Ec_{\beta }(\frac{%
\mathcal{E(-}t^{\beta })}{\hbar _{\beta }})+i^{\beta }Es_{\beta }(\frac{%
\mathcal{E(-}t^{\beta })}{\hbar _{\beta }})|^{2}\int\limits_{-\infty
}^{\infty }dx|\varphi (x)|^{2}.  \label{eq4sh-ener}
\end{equation}

With help of Eq.(\ref{eq42}) the last equation can be rewritten as

\begin{equation}
\mathrm{E}_{\beta }=\mathcal{E}\left( Ec_{\beta }^{2}(\frac{\mathcal{E(-}%
t^{\beta })}{\hbar _{\beta }})+Es_{\beta }^{2}(\frac{\mathcal{E(-}t^{\beta })%
}{\hbar _{\beta }})+2\cos \frac{\pi \beta }{2}Ec_{\beta }(\frac{\mathcal{E(-}%
t^{\beta })}{\hbar _{\beta }})Es_{\beta }(\frac{\mathcal{E(-}t^{\beta })}{%
\hbar _{\beta }})\right) .  \label{eq5sh-ener}
\end{equation}

This equation defines the energy $\mathrm{E}_{\beta }$ of time fractional
quantum system with the pseudo-Hamilton operator $\widehat{H}_{\alpha ,\beta
}(\widehat{p}_{\beta },\widehat{x})$ introduced by Eq.(\ref{eq13}). The
energy $\mathrm{E}_{\beta }$ defined by Eq.(\ref{eq3sh-ener}) is real due to
Hermiticity of the pseudo-Hamilton operator Eq.(\ref{eq13}). We see, that
the energy $\mathrm{E}_{\beta }$ of time fractional quantum system depends
on time $t$, eigenvalue $\mathcal{E}$ of pseudo-Hamilton operator, and
fractality parameter $\beta $. Thus, we come to the conclusion that in the
framework of time fractional quantum mechanics there are no stationary
states and the eigenvalues of the pseudo-Hamilton operator are not the
energy levels of time fractional quantum system.

Let us note, that in the limit case $\beta =1$ we have (see, Eqs.(\ref%
{eq14_c}) and (\ref{eq14_s}) in Appendix A),%
\begin{equation}
Ec_{\beta }(z)|_{\beta =1}=\cos (z),\qquad Es_{\beta }(z)|_{\beta =1}=\sin
(z),\qquad \cos \frac{\pi \beta }{2}|_{\beta =1}=0.  \label{eq52sh-ener}
\end{equation}

Then, it follows from Eqs.(\ref{eq5sh-ener}) and (\ref{eq52sh-ener}) that

\begin{equation}
\mathrm{E}_{\beta }|_{\beta =1}=\mathrm{E}_{1}=\mathcal{E},
\label{eq51sh-ener}
\end{equation}

here $\mathrm{E}_{1}$ is the energy of physical quantum system and $\mathcal{%
E}$ is the eigenvalue of quantum mechanical Hamilton operator $\widehat{H}%
_{\alpha }(\widehat{p},\widehat{x})$ given by Eq.(\ref{eq55}).

Therefore, in the limit case $\beta =1$ we recover the well-known statement
of standard quantum mechanics and fractional quantum mechanics \cite{Laskin1}%
-\cite{Laskin4} that the energy spectrum of a quantum system is a set of
eigenvalues of Hamilton operator.

\section{Applications of time fractional quantum mechanics}

\subsection{A free particle wave function}

For a free particle when $V(x,t)=0$, the 1D space-time fractional Schr\"{o}%
dinger equation (\ref{eq1}) reads

\begin{equation}
i^{\beta }\hbar _{\beta }\partial _{t}^{\beta }\psi (x,t)=D_{\alpha ,\beta
}(-\hbar _{\beta }^{2}\Delta )^{\alpha /2}\psi (x,t),\qquad 1<\alpha \leq
2,\qquad 0<\beta \leq 1,  \label{eq9sh_free}
\end{equation}

here $\psi (x,t)$ is the wave function, $i$\ is imanaginery unit, $i=\sqrt{-1%
}$, $\hbar _{\beta }$ is scale coefficient, $\Delta $ is 1D Laplace
operator, $\Delta =\partial ^{2}/\partial x^{2}$, and, finally, $\partial
_{t}^{\beta }$ is the left Caputo fractional derivative of order $\beta $
defined by Eq.(\ref{eq2}).

We are searching for solution to Eq.(\ref{eq9sh_free}) with the initial
condition $\psi _{0}(x)$,

\begin{equation}
\psi _{0}(x)=\psi (x,t=0).  \label{eq91}
\end{equation}

By applying the Fourier transform to the wave function $\psi (x,t)$

\begin{equation}
\psi (x,t)=\frac{1}{2\pi \hbar _{\beta }}\int\limits_{-\infty }^{\infty
}dp_{\beta }\exp \{ip_{\beta }x/\hbar _{\beta }\}\varphi (p_{\beta },t).
\label{eq92}
\end{equation}

with $\varphi (p_{\beta },t)$ defined by

\begin{equation}
\varphi (p_{\beta },t)=\int\limits_{-\infty }^{\infty }dx\exp \{-ip_{\beta
}x/\hbar _{\beta }\}\psi (x,t),  \label{eq93}
\end{equation}

we obtain from Eq.(\ref{eq9sh_free})

\begin{equation}
i^{\beta }\hbar _{\beta }\partial _{t}^{\beta }\varphi (p_{\beta
},t)=D_{\alpha ,\beta }|p_{\beta }^{2}|^{\alpha /2}\varphi (p_{\beta
},t)\qquad 1<\alpha \leq 2,\qquad 0<\beta \leq 1,  \label{eq12sh_free}
\end{equation}

with the initial condition $\varphi _{0}(p_{\beta })$ given by

\begin{equation}
\varphi _{0}(p_{\beta })=\varphi (p_{\beta },t=0)=\int\limits_{-\infty
}^{\infty }dx\exp \{-ip_{\beta }x/\hbar _{\beta }\}\psi _{0}(x).
\label{eq13sh_free}
\end{equation}

The equation (\ref{eq12sh_free}) is the 1D space-time fractional Schr\"{o}%
dinger equation in momentum representation.

The solution to the problem introduced by Eqs.(\ref{eq12sh_free}) and (\ref%
{eq13sh_free}) is

\begin{equation}
\varphi (p_{\beta },t)=E_{\beta }\left( D_{\alpha ,\beta }|p_{\beta
}^{2}|^{\alpha /2}\frac{t^{\beta }}{i^{\beta }\hbar _{\beta }}\right)
\varphi _{0}(p_{\beta }),  \label{eq14sh_free}
\end{equation}

where $E_{\beta }$ is the Mittag-Leffler function given by Eq.(\ref{eq52}).

Hence, the solution to the 1D space-time fractional Schr\"{o}dinger equation
Eq.(\ref{eq9sh_free}) with initial condition given by Eq.(\ref{eq91}) can be
presented as

\begin{equation}
\psi (x,t)=\frac{1}{2\pi \hbar _{\beta }}\int\limits_{-\infty }^{\infty
}dx^{\prime }\int\limits_{-\infty }^{\infty }dp_{\beta }\exp \{i\frac{%
p_{\beta }(x-x^{\prime })}{\hbar _{\beta }}\}E_{\beta }\left( D_{\alpha
,\beta }|p_{\beta }^{2}|^{\alpha /2}\frac{t^{\beta }}{i^{\beta }\hbar
_{\beta }}\right) \psi _{0}(x^{\prime }).  \label{eq15sh_free1}
\end{equation}

\subsection{A free particle space-time fractional quantum mechanical kernel}

The solution (\ref{eq15sh_free1}) to space-time fractional Schr\"{o}dinger
equation can be expressed in the form

\begin{equation}
\psi (x,t)=\int\limits_{-\infty }^{\infty }dx^{\prime }K_{\alpha ,\beta
}^{(0)}(x-x^{\prime },t)\psi _{0}(x^{\prime }),  \label{eq15sh_free2}
\end{equation}

if we introduce into consideration a free particle space-time fractional
quantum mechanical kernel $K_{\alpha ,\beta }^{(0)}(x,t)$ defined by

\begin{equation}
K_{\alpha ,\beta }^{(0)}(x,t)=\frac{1}{2\pi \hbar _{\beta }}%
\int\limits_{-\infty }^{\infty }dp_{\beta }\exp \{i\frac{p_{\beta }x}{\hbar
_{\beta }}\}E_{\beta }\left( D_{\alpha ,\beta }|p_{\beta }^{2}|^{\alpha /2}%
\frac{t^{\beta }}{i^{\beta }\hbar _{\beta }}\right) ,  \label{eq16sh_prop}
\end{equation}

\begin{equation*}
1<\alpha \leq 2,\qquad 0<\beta \leq 1.
\end{equation*}

The space-time fractional quantum mechanical kernel $K_{\alpha ,\beta
}^{(0)}(x,t)$ satisfies

\begin{equation}
K_{\alpha ,\beta }^{(0)}(x,t)|_{t=0}=K_{\alpha ,\beta }^{(0)}(x,0)=\delta
(x),\qquad 1<\alpha \leq 2,\qquad 0<\beta \leq 1,  \label{eq16sh_prop2}
\end{equation}

where $\delta (x)$ is delta function.

It follows immediately from Eq.(\ref{eq16sh_prop}) that the Fourier
transform $K_{\alpha ,\beta }^{(0)}(p_{\beta },t)$ of space-time fractional
quantum mechanical kernel is

\begin{equation}
K_{\alpha ,\beta }^{(0)}(p_{\beta },t)=\int\limits_{-\infty }^{\infty
}dx\exp \{-i\frac{p_{\beta }x}{\hbar _{\beta }}\}K_{\alpha ,\beta
}^{(0)}(x,t)  \label{eq17sh_prop}
\end{equation}

\begin{equation*}
=E_{\beta }\left( D_{\alpha ,\beta }|p_{\beta }^{2}|^{\alpha /2}\frac{%
t^{\beta }}{i^{\beta }\hbar _{\beta }}\right) ,\qquad 1<\alpha \leq 2,\qquad
0<\beta \leq 1.
\end{equation*}

This is space-time fractional quantum mechanical kernel in momentum
representation.

Applying the Laplace transform with respect to the time variable $t$ we
obtain the Fourier-Laplace transform $K_{\alpha ,\beta }^{(0)}(p_{\beta },u)$
of space-time fractional quantum mechanical kernel

\begin{equation}
K_{\alpha ,\beta }^{(0)}(p_{\beta },s)=\int\limits_{0}^{\infty }dt\exp
\{-st\}K_{\alpha ,\beta }^{(0)}(p_{\beta },t)  \label{eq18sh_propo}
\end{equation}

\begin{equation*}
=\frac{s^{\beta -1}}{s^{\beta }-D_{\alpha ,\beta }|p_{\beta }^{2}|^{\alpha
/2}/i^{\beta }\hbar _{\beta }}K_{\alpha ,\beta }^{(0)}(p_{\beta },0).
\end{equation*}

Taking into account Eq.(\ref{eq16sh_prop2}) we have

\begin{equation}
K_{\alpha ,\beta }^{(0)}(p_{\beta },s)=\frac{s^{\beta -1}}{s^{\beta
}-D_{\alpha ,\beta }|p_{\beta }^{2}|^{\alpha /2}/i^{\beta }\hbar _{\beta }}.
\label{eq18sh_prop}
\end{equation}

\subsubsection{Renormalization properties of the space-time fractional
quantum mechanical kernel}

Aiming\textit{\ }to study renormalization properties of the space-time
fractional quantum mechanical kernel, let us re-write $K_{\alpha ,\beta
}^{(0)}(p_{\beta },s)$ \ given by Eq.(\ref{eq18sh_prop}) in integral form

\begin{equation}
K_{\alpha ,\beta }^{(0)}(p_{\beta },s)=s^{\beta -1}\int\limits_{0}^{\infty
}du\exp \{-u(s^{\beta }-D_{\alpha ,\beta }|p_{\beta }^{2}|^{\alpha
/2}/i^{\beta }\hbar _{\beta })\}  \label{eq20sh_prop}
\end{equation}

\begin{equation*}
=\int\limits_{0}^{\infty }duN_{\alpha ,\beta }(p_{\beta },u)L_{\beta }(u,s),
\end{equation*}

where we introduced two functions $N_{\alpha ,\beta }(p_{\beta },u)$ and $%
L_{\beta }(u,s)$ defined by

\begin{equation}
N_{\alpha ,\beta }(p_{\beta },u)=\exp \{uD_{\alpha ,\beta }|p_{\beta
}^{2}|^{\alpha /2}/i^{\beta }\hbar _{\beta }\},  \label{eq21sh_prop}
\end{equation}

and

\begin{equation}
L_{\beta }(u,s)=s^{\beta -1}\exp \{-us^{\beta }\}.  \label{eq22sh_prop}
\end{equation}

Having new function $N_{\alpha ,\beta }(p_{\beta },u)$ we obtain

\begin{equation}
N_{\alpha ,\beta }(x,u)=\frac{1}{2\pi \hbar _{\beta }}\int\limits_{-\infty
}^{\infty }dp_{\beta }\exp \{i\frac{p_{\beta }x}{\hbar _{\beta }}\}N_{\alpha
,\beta }(p_{\beta },u)  \label{eq23sh_prop}
\end{equation}

\begin{equation*}
=\frac{1}{2\pi \hbar _{\beta }}\int\limits_{-\infty }^{\infty }dp_{\beta
}\exp \{i\frac{p_{\beta }x}{\hbar _{\beta }}\}\exp \{uD_{\alpha ,\beta
}|p_{\beta }^{2}|^{\alpha /2}/i^{\beta }\hbar _{\beta }\}.
\end{equation*}

It is easy to see, that the scaling

\begin{equation}
N_{\alpha ,\beta }(x,u)=\frac{1}{u^{1/\alpha }}N_{\alpha ,\beta }(\frac{x}{%
u^{1/\alpha }},1),  \label{eq24sh_old}
\end{equation}

holds for function $N_{\alpha ,\beta }(x,u)$ given by Eq.(\ref{eq23sh_prop}).

Having new function $L_{\beta }(u,s)$ we obtain

\begin{equation}
L_{\beta }(u,t)=\frac{1}{2\pi i}\int\limits_{\sigma -i\infty }^{\sigma
+i\infty }dse^{st}L_{\beta }(u,s)  \label{eq24sh_prop}
\end{equation}

\begin{equation*}
=\frac{1}{2\pi i}\int\limits_{\sigma -i\infty }^{\sigma +i\infty
}dse^{st}s^{\beta -1}\exp \{-us^{\beta }\},
\end{equation*}

where $\sigma $ is greater than the real part of all singularities in the
complex plane of $L_{\beta }(u,s)$.

It is easy to see that the scaling

\begin{equation}
L_{\beta }(u,t)=\frac{1}{t^{\beta }}L_{\beta }(\frac{u}{t^{\beta }},1),
\label{eq241sh_prop}
\end{equation}

holds for function $L_{\beta }(u,t)$ given by Eq.(\ref{eq24sh_prop}).

In terms of above introduced functions $N_{\alpha ,\beta }(x,u)$ and $%
L_{\beta }(u,t)$ Eq.(\ref{eq16sh_prop}) reads

\begin{equation}
K_{\alpha ,\beta }^{(0)}(x,t)=\int\limits_{0}^{\infty }duN_{\alpha ,\beta
}(x,u)L_{\beta }(u,t),  \label{eq25sh_prop}
\end{equation}

where functions $N_{\alpha ,\beta }(x,u)$ and $L_{\beta }(u,t)$ are defined
by Eqs.(\ref{eq23sh_prop}) and (\ref{eq24sh_prop}).

The equivalent representation for$K_{\alpha ,\beta }^{(0)}(x,t)$ is

\begin{equation}
K_{\alpha ,\beta }^{(0)}(x,t)=\frac{1}{t^{\beta }}\int\limits_{0}^{\infty }du%
\frac{1}{u^{1/\alpha }}N_{\alpha ,\beta }(\frac{x}{u^{1/\alpha }},1)L_{\beta
}(\frac{u}{t^{\beta }},1),  \label{eq251sh_prop}
\end{equation}

\begin{equation*}
1<\alpha \leq 2,\qquad 0<\beta \leq 1.
\end{equation*}

Hence, one can consider Eq.(\ref{eq251sh_prop}) as an alternative
representation for space-time fractional quantum mechanical kernel $%
K_{\alpha ,\beta }^{(0)}(x,t)$. The interesting feature of this
representation is that it separates space and time variables for space-time
fractional quantum mechanical kernel $K_{\alpha ,\beta }^{(0)}(x,t)$. It
follows from Eqs.(\ref{eq23sh_prop}) and (\ref{eq25sh_prop}) that

\begin{equation}
K_{\alpha ,\beta }^{(0)}(x,t)=K_{\alpha ,\beta }^{(0)}(-x,t).
\label{eq26sh_prop}
\end{equation}

With help of Eqs.(\ref{eq24sh_old}) and (\ref{eq241sh_prop}) we conclude
that the scaling

\begin{equation}
K_{\alpha ,\beta }^{(0)}(x,t)=\frac{1}{t^{\beta /\alpha }}K_{\alpha ,\beta
}^{(0)}(\frac{x}{t^{\beta /\alpha }},1),  \label{eq27sh_prop}
\end{equation}

holds for space-time fractional quantum mechanical kernel $K_{\alpha ,\beta
}^{(0)}(x,t)$ given by Eq.(\ref{eq16sh_prop}).

Defined by Eqs.(\ref{eq241sh_prop}) and (\ref{eq24sh_prop}) function $%
L_{\beta }(u/t^{\beta },1)$ can be re-written as

\begin{equation}
L_{\beta }(z,1)=\frac{1}{2\pi i}\int\limits_{\sigma -i\infty }^{\sigma
+i\infty }dse^{s}s^{\beta -1}\exp \{-zs^{\beta }\}\qquad 0<\beta \leq 1,
\label{eq28sh_prop}
\end{equation}

where we introduced a new variable $z=u/t^{\beta }$.

\subsubsection{Wright function in time fractional quantum mechanics}

Our intent now is to show that the function $L_{\beta }(z,1)$ defined by Eq.(%
\ref{eq28sh_prop}) can be presented in terms of the Wright function (see,
for example, \cite{Bateman3}). By deforming the integration path in the
right hand side of Eq.(\ref{eq28sh_prop}) into the Hankel contour $H_{a}$ we
have

\begin{equation}
L_{\beta }(z,1)=\frac{1}{2\pi i}\int\limits_{H_{a}}^{{}}dse^{s}s^{\beta
-1}\exp \{-zs^{\beta }\},\qquad 0<\beta \leq 1.  \label{eq29_L}
\end{equation}

A series expansion for $\exp \{-zs^{\beta }\}$ yields

\begin{equation*}
L_{\beta }(z,1)=\frac{1}{2\pi i}\int\limits_{H_{a}}^{{}}dse^{s}s^{\beta
-1}\sum\limits_{k=0}^{\infty }\frac{(-1)^{k}}{k!}s^{\beta k}
\end{equation*}

\begin{equation}
=\frac{1}{2\pi i}\int\limits_{H_{a}}^{{}}dse^{s}s^{\beta
-1}\sum\limits_{k=0}^{\infty }\frac{(-1)^{k}z^{k}}{k!}s^{\beta k}
\label{eq30_L}
\end{equation}

\begin{equation*}
=\sum\limits_{k=0}^{\infty }\frac{(-1)^{k}z^{k}}{k!}\left( \frac{1}{2\pi i}%
\int\limits_{H_{a}}^{{}}dse^{s}s^{\beta k+\beta -1}\right) ,\qquad 0<\beta
\leq 1.
\end{equation*}

Using the well-known Hankel representation of the Gamma function (see, for
example, \cite{Bateman1}),

\begin{equation}
\frac{1}{\Gamma (-\sigma k+\nu )}=\left( \frac{1}{2\pi i}\int%
\limits_{H_{a}}^{{}}dse^{s}s^{\sigma k-\nu }\right) ,\qquad 0<\beta \leq 1,
\label{eq31_L}
\end{equation}

we obtain for $L_{\beta }(z,1)$

\begin{equation}
L_{\beta }(z,1)=\sum\limits_{k=0}^{\infty }\frac{(-1)^{k}z^{k}}{k!\Gamma
(-\beta k+(1-\beta ))}.  \label{eq32_1}
\end{equation}

In the notations of Ref.\cite{Bateman3}, the Wright function $\phi (a,\beta
;z)$ is expressed as

\begin{equation}
\phi (a,\beta ;z)=\sum\limits_{k=0}^{\infty }\frac{z^{k}}{k!\Gamma (ak+\beta
))},\qquad -1<a<0,\qquad \beta \geq 0.  \label{eq33_L}
\end{equation}

Therefore, we come to following representation of our function $L_{\beta
}(z,1)$ in terms of the Wright function $\phi $

\begin{equation}
L_{\beta }(z,1)=\phi (-\beta ,1-\beta ;-z).  \label{eq34_L}
\end{equation}

Knowledge of particular cases for the Wright function let us obtain the
particular cases for the function $L_{\beta }(z,1)$.

1. When $\beta =1/2$ we have

\begin{equation}
L_{1/2}(z,1)=L_{\beta }(z,1)|_{\beta =1/2}=\frac{1}{\sqrt{\pi }}\exp (-\frac{%
z^{2}}{4}).  \label{eq341_L}
\end{equation}

2. When $\beta =1/2$ we have

\begin{equation}
L_{1}(z,1)=L_{\beta }(z,1)|_{\beta =1}=\delta (z-1),  \label{eq35_L}
\end{equation}

where $\delta (z)$ is delta function. This property is useful to recover
fractional quantum mechanics\ and standard quantum mechanics formulas from
the general equation (\ref{eq25sh_prop}).

Thus, function $L_{\beta }(z,1)$ involved into representation (\ref%
{eq251sh_prop}) of space-time fractional quantum kernel, is expressed by Eq.(%
\ref{eq34_L}) as the Wright function. It introduces into the time fractional
quantum mechanics a well developed mathematical tool - Wright function. In
other words, many well-known results related to the Wright function can be
applied to study fundamental properties of the space-time fractional quantum
kernel and develop a variety of new applications of time fractional quantum
mechanics.

\subsubsection{Fox \textit{H}-function representation for a free particle
space-time fractional quantum kernel}

The space-time fractional quantum kernel given by Eq.(\ref{eq16sh_prop}) can
be expressed in terms of the Fox $H$-function. The definition of the $H$%
-function and its fundamental properties can be found in \cite{Mathai1}. For
paper integrity and reader's audience convenience we presented the
definition of the Fox $H$-function and some its properties in Appendix B.

In terms of the Fox $H$-function the Mittag-Leffler function $E_{\beta }(z)$
is presented by Eq.(\ref{eqA.5}), see, Appendix B. Then, $K_{\alpha ,\beta
}^{(0)}(x,t)$ given by Eq.(\ref{eq16sh_prop}) reads

\begin{equation}
K_{\alpha ,\beta }^{(0)}(x,t)=\frac{1}{\pi \hbar _{\beta }}%
\int\limits_{0}^{\infty }dp_{\beta }\cos (\frac{p_{\beta }x}{\hbar _{\beta }}%
)H_{1,2}^{1,1}\left[ -D_{\alpha ,\beta }|p^{2}|^{\alpha /2}\frac{t^{\beta }}{%
i^{\beta }\hbar _{\beta }}|\QATOP{(0,1)}{(0,1),(0,\beta )}\right] .
\label{eq16_free_3}
\end{equation}

With help of the cosine transform of the $H$-function defined by of Eq.(\ref%
{eqB_1a}), see Appendix B, we obtain

\begin{equation}
K_{\alpha ,\beta }^{(0)}(x,t)=\frac{1}{|x|}H_{3,3}^{2,1}\left[ -\frac{%
i^{\beta }\hbar _{\beta }}{D_{\alpha ,\beta }\hbar _{\beta }^{\alpha
}t^{\beta }}|x|^{\alpha }|\QATOP{(1,1),(1,\beta ),(1,\alpha /2)}{(1,\alpha
),(1,1),(1,\alpha /2)}\right] ,  \label{eq16_free_41}
\end{equation}

\begin{equation*}
1<\alpha \leq 2,\qquad 0<\beta \leq 1,
\end{equation*}

which is the expression for the space-time fractional quantum kernel $%
K_{\alpha ,\beta }^{(0)}(x,t)$ in terms of $H_{3,3}^{2,1}$-function.

Alternatively, using the Property 4 given by Eq.(\ref{eqA.7}), see Appendix
B, $K_{\alpha ,\beta }^{(0)}(x,t)$ can be presented as

\begin{equation}
K_{\alpha ,\beta }^{(0)}(x,t)=\frac{1}{\alpha |x|}H_{3,3}^{2,1}\left[ \frac{1%
}{\hbar _{\beta }}\left( -\frac{i^{\beta }\hbar _{\beta }}{D_{\alpha ,\beta
}t^{\beta }}\right) ^{1/\alpha }|x||\QATOP{(1,1/\alpha ),(1,\beta /\alpha
),(1,1/2)}{(1,1),(1,1/\alpha ),(1,1/2)}\right] ,  \label{eq17_free}
\end{equation}

\begin{equation*}
1<\alpha \leq 2,\qquad 0<\beta \leq 1.
\end{equation*}

Thus, the equations (\ref{eq16_free_41}) and \ (\ref{eq17_free}) introduce a
new family of a free particle space-time fractional quantum mechanical
kernels parametrized by two fractality parameters $\alpha $ and $\beta $.
Newly introduced kernel $K_{\alpha ,\beta }^{(0)}(x,\tau )$ does not satisfy
quantum superposition law defined by Eq.(24) in \cite{Laskin2}, due to the
fractional time derivative of order $\beta $ in Eq.(\ref{eq9sh_free}).

\section{Special cases \ of time fractional quantum mechanics}

With particular choice of fractality parameters $\alpha $ and $\beta $ the
space-time fractional Schr\"{o}dinger equation Eq.(\ref{eq1}) covers the
following three special cases: the Schr\"{o}dinger equation ($\alpha =2$\
and $\beta =1$), the fractional Schr\"{o}dinger equation ($1<\alpha \leq 2$\
and $\beta =1$) and the time fractional Schr\"{o}dinger equation ($\alpha =2$
and $0<\beta \leq 1$).

\subsection{The Schr\"{o}dinger equation - the case when $\protect\alpha =2$
and $\protect\beta =1$}

When $\alpha =2$ and $\beta =1$ we have

\begin{equation}
D_{\alpha ,\beta }|_{\alpha =2,\beta =1}=D_{2,1}=\frac{1}{2m},\qquad
p_{\beta }|_{\beta =1}=p,  \label{eq40_1}
\end{equation}

and%
\begin{equation}
\hbar _{\beta }|_{\beta =1}=\hbar ,  \label{eq40_2}
\end{equation}

where $m$ is mass of a quantum particle, $p$ is momentum of a quantum
particle and $\hbar $ is the well-known Planck's constant.

In this case space-time fractional Schr\"{o}dinger equation Eq.(\ref{eq1})
goes into the celebrated Schr\"{o}dinger equation \cite{Schrodinger}

\begin{equation}
i\hbar \partial _{t}\psi (x,t)=-\frac{\hbar ^{2}}{2m}\Delta \psi
(x,t)+V(x,t)\psi (x,t).  \label{eq40_3}
\end{equation}

For the quantum mechanical kernel $K_{2,1}^{(0)}(x,t)=$ $K_{\alpha ,\beta
}^{(0)}(x,t)|_{\alpha =2,\beta =1},$ where $K_{\alpha ,\beta }^{(0)}(x,t)$
is defined by Eq.(\ref{eq25sh_prop}) we obtain

\begin{equation}
K_{2,1}^{(0)}(x,t)=\int\limits_{0}^{\infty }duN_{2,1}(x,u)L_{1}(u,t),
\label{eq40_4}
\end{equation}

here

\begin{equation}
N_{2,1}(x,u)=\frac{1}{2\pi \hbar }\int\limits_{-\infty }^{\infty }dp\exp \{i%
\frac{px}{\hbar }\}\exp \{uD_{2,1}p^{2}/i\hbar \}.  \label{eq40_5}
\end{equation}

\begin{equation*}
=\frac{1}{2\pi \hbar }\int\limits_{-\infty }^{\infty }dp\exp \{i\frac{px}{%
\hbar }\}\exp \{-iu\frac{p^{2}}{2m\hbar }\},
\end{equation*}

and

\begin{equation}
L_{1}(u,t)=\frac{1}{2\pi i}\int\limits_{\sigma -i\infty }^{\sigma +i\infty
}dse^{st}\exp \{-us\}=\delta (t-u).  \label{eq40_6}
\end{equation}

Substitution of Eqs.(\ref{eq40_5}) and (\ref{eq40_6}) into Eq.(\ref{eq40_4})
yields

\begin{equation}
K_{2,1}^{(0)}(x,t)=\frac{1}{2\pi \hbar }\int\limits_{-\infty }^{\infty
}dp\exp \{i\frac{px}{\hbar }\}\exp \{-it\frac{p^{2}}{2m\hbar }\}.
\label{eq40_7}
\end{equation}

The intergral in Eq.(\ref{eq40_7}) can be evaluated analytically and the
result is

\begin{equation}
K_{2,1}^{(0)}(x,t)=\sqrt{\frac{m}{2\pi i\hbar t}}\exp \{i\frac{mx^{2}}{%
2\hbar t}\},  \label{eq40_8}
\end{equation}

which is Feynman's free particle quantum kernel \cite{Feynman}.

\subsection{The fractional Schr\"{o}dinger equation - the case when $1<%
\protect\alpha \leq 2$ and $\protect\beta =1$}

In the case when $\beta =1$, $D_{\alpha ,\beta }$ goes into the scale
coefficient $D_{\alpha }$ emerging in fractional quantum mechanics \cite%
{Laskin1}, \cite{Laskin2},

\begin{equation}
D_{\alpha ,\beta }|_{\beta =1}=D_{\alpha ,1}=D_{\alpha },\qquad 1<\alpha
\leq 2,  \label{eq24dim}
\end{equation}

which has physical dimension,

\begin{equation}
\lbrack D_{\alpha }]=\mathrm{erg}^{1-\alpha }\cdot \mathrm{cm}^{\alpha
}\cdot \mathrm{sec}^{-\alpha }.  \label{eq25dim}
\end{equation}

In this case the space-time fractional Schr\"{o}dinger equation Eq.(\ref{eq1}%
) goes into Laskin's fractional Schr\"{o}dinger equation \cite{Laskin4}

\begin{equation}
i\hbar \partial _{t}\psi (x,t)=D_{\alpha }(-\hbar ^{2}\Delta )^{\alpha
/2}\psi (x,t)+V(x,t)\psi (x,t),\qquad 1<\alpha \leq 2,  \label{eq26sh}
\end{equation}

with $\hbar $\ being fundamental Planck's constant.

For the quantum mechanical kernel $K_{\alpha ,1}^{(0)}(x,t)=$ $K_{\alpha
,\beta }^{(0)}(x,t)|_{\beta =1},$ where $K_{\alpha ,\beta }^{(0)}(x,t)$ is
defined by Eq.(\ref{eq25sh_prop}) we obtain

\begin{equation}
K_{\alpha ,1}^{(0)}(x,t)=\int\limits_{0}^{\infty }duN_{\alpha
,1}(x,u)L_{1}(u,t),  \label{eq40_4a}
\end{equation}

here

\begin{equation}
N_{\alpha ,1}(x,u)=\frac{1}{2\pi \hbar }\int\limits_{-\infty }^{\infty
}dp\exp \{i\frac{px}{\hbar }\}\exp \{uD_{\alpha }|p^{2}|^{\alpha /2}/i\hbar
\},  \label{eq40_5a}
\end{equation}

and

\begin{equation}
L_{1}(u,t)=\frac{1}{2\pi i}\int\limits_{\sigma -i\infty }^{\sigma +i\infty
}dse^{st}\exp \{-us\}=\delta (t-u).  \label{eq40_6a}
\end{equation}

Substitution of Eqs.(\ref{eq40_5a}) and (\ref{eq40_6a}) into Eq.(\ref%
{eq40_4a}) yields

\begin{equation}
K_{\alpha ,1}^{(0)}(x,t)=\frac{1}{2\pi \hbar }\int\limits_{-\infty }^{\infty
}dp\exp \{i\frac{px}{\hbar }\}\exp \{-it\frac{D_{\alpha }|p^{2}|^{\alpha /2}%
}{\hbar }\}.  \label{eq40_7a}
\end{equation}

It was shown in \cite{Laskin2}, \cite{Laskin5} that the integral in Eq.(\ref%
{eq40_7a}) can be expressed in terms of the $H_{2,2}^{1,1}$-function,

\begin{equation}
K_{\alpha ,1}^{(0)}(x,t)=\frac{1}{\alpha |x|}H_{2,2}^{1,1}\left[ \frac{1}{%
\hbar }\left( -\frac{i\hbar }{D_{\alpha }t}\right) ^{1/\alpha }|x||\QATOP{%
(1,1/\alpha ),(1,1/2)}{(1,1),(1,1/2)}\right] ,\qquad 1<\alpha \leq 2.
\label{eq40_7b}
\end{equation}

The expression (\ref{eq40_7b}) for a free particle quantum kernel $K_{\alpha
,1}^{(0)}(x,t)$ can be obtained directly from Eq.(\ref{eq17_free}). It is
easy to see, that when $\beta =1$ Eq.(\ref{eq17_free}) reads

\begin{equation*}
K_{\alpha ,1}^{(0)}(x,t)=K_{\alpha ,\beta }^{(0)}(x,t)|_{\beta =1}
\end{equation*}

\begin{equation}
=\frac{1}{\alpha |x|}H_{3,3}^{2,1}\left[ \frac{1}{\hbar }\left( -\frac{%
i\hbar }{D_{\alpha }t}\right) ^{1/\alpha }|x||\QATOP{(1,1/\alpha
),(1,1/\alpha ),(1,1/2)}{(1,1),(1,1/\alpha ),(1,1/2)}\right] ,
\label{eq27sh_1}
\end{equation}

\begin{equation*}
1<\alpha \leq 2.
\end{equation*}

By using Property 1 (see, Appendix B) we re-write the above expression as

\begin{equation}
K_{\alpha ,1}^{(0)}(x,t)=\frac{1}{\alpha |x|}H_{3,3}^{2,1}\left[ \frac{1}{%
\hbar }\left( -\frac{i\hbar }{D_{\alpha }t}\right) ^{1/\alpha }|x||\QATOP{%
(1,1/\alpha ),(1,1/2),(1,1/\alpha )}{(1,1/\alpha ),(1,1),(1,1/2)}\right] .
\label{eq27sh_2}
\end{equation}

The next step is to use Property 2 (see, Appendix B) to come to Eq.(\ref%
{eq40_7a}) for a free particle fractional quantum kernel $K_{\alpha
,1}^{(0)}(x,t)$. It has been shown in \cite{Laskin5} that at $\alpha =2$ Eq.(%
\ref{eq40_7b}) goes into Eq.(\ref{eq40_8}).

\subsection{The time Schr\"{o}dinger equation - the case when $\protect%
\alpha =2$ and $0<\protect\beta \leq 1$}

In the case when $\alpha =2$ we have

\begin{equation}
D_{\alpha ,\beta }|_{\alpha =2}=D_{2,\beta },\qquad 0<\beta \leq 1,
\label{eq22dim}
\end{equation}

where $D_{2,\beta }$ is the scale coefficient with physical dimension $%
[D_{2,\beta }]=\mathrm{g}^{-1}\mathrm{\cdot sec}^{2-2\beta }$.

In this case the space-time fractional Schr\"{o}dinger equation Eq.(\ref{eq1}%
) goes into the time fractional Schr\"{o}dinger equation of the form

\begin{equation}
i^{\beta }\hbar _{\beta }\partial _{t}^{\beta }\psi (x,t)=-D_{2,\beta }\hbar
_{\beta }^{2}\Delta \psi (x,t)V(x,t)\psi (x,t).,\qquad 0<\beta \leq 1.
\label{eq23_Laskin}
\end{equation}

This equation can be considered as an alternative to Naber's time fractional
Schr\"{o}dinger equation \cite{Naber}.

For the quantum mechanical kernel $K_{2,\beta }^{(0)}(x,t)=$ $K_{\alpha
,\beta }^{(0)}(x,t)|_{\alpha =2},$ where $K_{\alpha ,\beta }^{(0)}(x,t)$ is
defined by Eq.(\ref{eq25sh_prop}) we obtain

\begin{equation}
K_{2,\beta }^{(0)}(x,t)=\int\limits_{0}^{\infty }duN_{2,\beta }(x,u)L_{\beta
}(u,t),  \label{eq40_4a}
\end{equation}

here

\begin{equation}
N_{2,\beta }(x,u)=\frac{1}{2\pi \hbar _{\beta }}\int\limits_{-\infty
}^{\infty }dp_{\beta }\exp \{i\frac{p_{\beta }x}{\hbar }\}\exp \{uD_{2,\beta
}p_{\beta }^{2}/i^{\beta }\hbar _{\beta }\},  \label{eq40_5a}
\end{equation}

and $L_{\beta }(u,t)$ is given by Eq.(\ref{eq24sh_prop}).

\begin{equation}
L_{\beta }(u,t)=\frac{1}{2\pi i}\int\limits_{\sigma -i\infty }^{\sigma
+i\infty }dse^{st}L_{\beta }(u,s)
\end{equation}

\begin{equation*}
=\frac{1}{2\pi i}\int\limits_{\sigma -i\infty }^{\sigma +i\infty
}dse^{st}s^{\beta -1}\exp \{-us^{\beta }\}.
\end{equation*}

To express a free particle time fractional quantum kernel $K_{2,\beta
}^{(0)}(x,t)$ in terms of the $H$-function let's put $\alpha =2$ in Eq.(\ref%
{eq17_free})

\begin{equation*}
K_{2,\beta }^{(0)}(x,\tau )=K_{\alpha ,\beta }^{(0)}(x,\tau )|_{\alpha =2}
\end{equation*}

\begin{equation}
=\frac{1}{2|x|}H_{3,3}^{2,1}\left[ \frac{1}{\hbar _{\beta }}\left( -\frac{%
i^{\beta }\hbar _{\beta }}{D_{\alpha ,\beta }\tau ^{\beta }}\right) ^{1/2}|x|%
{\LARGE |}\QATOP{(1.1/2),(1,\beta /2),(1,1/2)}{(1,1),(1,1/2),(1,1/2)}\right]
,  \label{eq23sh_1}
\end{equation}

\begin{equation*}
0<\beta \leq 1.
\end{equation*}

Using the Property 2 yields

\begin{equation}
K_{2,\beta }^{(0)}(x,\tau )=\frac{1}{2|x|}H_{2,2}^{2,0}\left[ \frac{1}{\hbar
_{\beta }}\left( -\frac{i^{\beta }\hbar _{\beta }}{D_{\alpha ,\beta }\tau
^{\beta }}\right) ^{1/2}|x|{\LARGE |}\QATOP{(1,\beta /2),(1,1/2)}{%
(1,1),(1,1/2)}\right] .
\end{equation}

Next, by using the Property 1 and then the Property 2 we find

\begin{equation}
K_{2,\beta }^{(0)}(x,\tau )=\frac{1}{2|x|}H_{1,1}^{1,0}\left[ \frac{1}{\hbar
_{\beta }}\left( -\frac{i^{\beta }\hbar _{\beta }}{D_{\alpha ,\beta }\tau
^{\beta }}\right) ^{1/2}|x|{\LARGE |}\QATOP{(1,\beta /2)}{(1,1)}\right] .
\label{eq25sh_prop_beta}
\end{equation}

Thus, we found a new expression of a free particle time fractional quantum
kernel $K_{2,\beta }^{(0)}(x,\tau )$ in terms of $H_{1,1}^{1,0}$-finction.
It can be shown that at $\beta =1$ Eq.(\ref{eq25sh_prop_beta}) goes into Eq.(%
\ref{eq40_8}).

\section{Conclusion}

Fractional quantum mechanics has emerged as a field over 15 years ago,
attracting the attention of many researchers. The original idea behind
fractional quantum mechanics was to develop a path integral over L\'{e}%
vy-like quantum paths instead of the well-known Feynman path integral over
Brownian-like quantum paths. The basic outcome of implementing this idea is
an alternative path integral approach, which results in a new fundamental
equation -- the fractional Schr\"{o}dinger equation. In other words, if the
Feynman path integral approach to quantum mechanics allows one to reproduce
the Schr\"{o}dinger equation, then the path integral over Levy-like paths
leads one to the fractional Schr\"{o}dinger equation. This is a
manifestation of a new non-Gaussian paradigm, based on deep relationships
between the structure of fundamental physics equations and fractal
dimensions of \textquotedblleft underlying\textquotedblright\ quantum paths.
The fractional Schr\"{o}dinger equation includes the spatial derivative of
fractional order $\alpha $ ($\alpha $ is the L\'{e}vy index), instead of the
second order spatial derivative in the well-known Schr\"{o}dinger equation.
Thus, the fractional Schr\"{o}dinger equation is the fractional differential
equation in accordance with modern terminology. This is the main point of
the term, \textit{fractional Schr\"{o}dinger equation}, and the more general
term,\ \textit{fractional quantum mechanics}.

In this paper we introduce the concept of time fractional quantum mechanics.
The wording "\textit{time fractional quantum mechanics}" means that the time
derivative in the fundamental quantum mechanical equations - Schr\"{o}dinger
equation and fractional Schr\"{o}dinger equation, is substituted with a
fractional time derivative. The time fractional derivative in our approach
is the Caputo fractional derivative.

It has to be noted, that despite some shortcomings from the stand point of
quantum mechanical fundamentals, time fractional quantum mechanics attracts
the attention of researchers as an interesting new application of fractional
calculus to quantum mechanics. Many preliminary developments were done in
the field of time fractional quantum mechanics. Naber invented \textit{time
fractional Schr\"{o}dinger equation }\cite{Naber}. The time fractional Schr%
\"{o}dinger equation involves a time derivative of fractional order instead
of first-order time derivative, while the spatial derivative is the
second-order spatial derivative as it is in the well-known Schr\"{o}dinger
equation. Naber has found the exact solutions to the time fractional Schr%
\"{o}dinger equation for a free particle and a particle in a potential well 
\cite{Naber}.

Later on, Wang and Xu \cite{WangandXu}, and then Dong and Xu \cite{DongandXu}%
, combined both Laskin's equation and Naber's equation and came up with 
\textit{space-time fractional Schr\"{o}dinger equation}. The space-time
fractional Schr\"{o}dinger equation includes both spatial and temporal
fractional derivatives. Wang and Xu found exact solutions to space-time
fractional Schr\"{o}dinger equation for a free particle and for an infinite
square potential well \cite{WangandXu}. Dong and Xu found the exact solution
to space-time fractional Schr\"{o}dinger equation for a quantum particle in $%
\delta $-potential well \cite{DongandXu}.

To introduce and develop a time fractional quantum mechanics we begin with
our own version of the space-time fractional Schr\"{o}dinger equation. Our
space-time fractional Schr\"{o}dinger equation involves two scale
dimensional parameters, one of which can be considered as time fractional
generalization of the famous Planck's constant, while the other one can be
interpreted as a time fractional generalization of the scale parameter
introduced by Laskin in fractional quantum mechanics \cite{Laskin1}-\cite%
{Laskin4}. The fractional generalization of Planck's constant is a
fundamental dimensional parameter of time fractional quantum mechanics,
while the time fractional generalization of Laskin's scale parameter \cite%
{Laskin1}-\cite{Laskin4} plays a fundamental role in both time fractional
quantum mechanics and time fractional classical mechanics.

In addition to the above mentioned dimensional parameters, time fractional
quantum \ mechanics involves two dimensionless fractality parameters $\alpha 
$, $1<\alpha \leq 2$ and.$\beta $, $0<\beta \leq 1$. Parameter $\alpha $ is
the order of spatial fractional quantum Riesz derivative \cite{Laskin1}, 
\cite{Laskin2} and $\beta $ is the order of time fractional derivative
Caputo derivative. In other words, $\alpha $ is responsible for modelling 
\textit{spatial fractality}, while the parameter $\beta $, is responsible
for modeling \textit{temporal fractality}.

Time fractional quantum mechanical operators of coordinate, momentum and
angular momentum have been introduced and their commutation relationship has
been established. The pseudo-Hamilton \ quantum mechanical operator has been
introduced and its Hermiticity has been proven in the framework of time
fractional quantum mechanics. The general solution to the space-time
fractional Schr\"{o}dinger equation was found in the case when the
pseudo-Hamilton operator does not depend on time. Energy of a quantum system
in the framework of time fractional quantum mechanics was defined and
calculated in terms of the Mittag-Leffler function. Two new functions
associated with the Mittag-Leffler function have been launched and
elaborated. These two new functions can be considered as a natural
fractional generalization of the well-known trigonometric functions sine and
cosine. A fractional generalization of the celebrated Euler equation was
discovered. A free particle space-time fractional quantum kernel was
calculated in terms of the Fox's $H$-functions. It has been shown that a
free particle space-time fractional quantum kernel can be alternatively
expressed with help of the Wright function.

The framework of time fractional quantum mechanics, depending on choices of
fractality parameters $\alpha $ and $\beta $, covers the following
fundamental quantum equations:

1. The Schr\"{o}dinger equation (Schr\"{o}dinger equation \cite{Schrodinger}%
), $\alpha =2$\ and $\beta =1$;

2. Fractional Schr\"{o}dinger equation (Laskin equation \cite{Laskin4}), $%
1<\alpha \leq 2$\ and $\beta =1$;

3. Time fractional Schr\"{o}dinger equation (Naber equation \cite{Naber}), \ 
$\alpha =2$ and $0<\beta \leq 1$;

4. Space-time fractional Schr\"{o}dinger equation\ (Wang and Xu \cite%
{WangandXu} and Dong and Xu \cite{DongandXu} equation), $1<\alpha \leq 2$\
and $0<\beta \leq 1$.

\section{Acknowledgments}

I am grateful to Prof. Mark M. Meerschaert, for inviting me to give a talk
at \textit{A Workshop on Future Directions in Fractional Calculus Research
and Applications}, which took place on 17 - 21 October 2016, at Michigan
State University, East Lansing, MI 48823.

\section{Appendix A}

\subsection{Fractional generalization of trigonometric functions $\cos (z)$
and $\sin (z)$}

Here we introduce two new functions $Ec_{\beta }(z)$ and $Es_{\beta }(z)$
defined by the series

\begin{equation}
Ec_{\beta }(z)=\sum\limits_{m=0}^{\infty }\frac{(-1)^{m\beta }z^{2m}}{\Gamma
(2\beta m+1)},  \label{eq13_sol}
\end{equation}

and

\begin{equation}
Es_{\beta }(z)=\sum\limits_{m=0}^{\infty }\frac{(-1)^{m\beta }z^{(2m+1)}}{%
\Gamma (\beta (2m+1)+1)}.  \label{eq14_sol}
\end{equation}

From definitions (\ref{eq13_sol})-(\ref{eq14_sol}) we can obtain,

\begin{equation}
Ec_{\beta }(z)=\frac{i^{\beta }E_{\beta }(-i^{\beta }z)-(-i)^{\beta
}E_{\beta }(i^{\beta }z)}{i^{\beta }-(-i)^{\beta }},  \label{eq15_sol1new}
\end{equation}

and

\begin{equation}
Es_{\beta }(z)=\frac{E_{\beta }(i^{\beta }z)-E_{\beta }(-i^{\beta }z)}{%
i^{\beta }-(-i)^{\beta }},  \label{eq15_sol2new}
\end{equation}

where $E_{\beta }(z)$ is the Mittag-Leffler function given by Eq.(\ref{eq52}%
) and $i$\ is imanaginery unit, $i=\sqrt{-1}$.

Hence, functions $Ec_{\beta }(z)$ and $Es_{\beta }(z)$ can be considered as
a natural fractional generalization of the well-known trigonometric
functions $\cos (z)$, and $\sin (z)$ respectively. Indeed, when $\beta =1$ $%
Ec_{\beta }(z)$ and $Es_{\beta }(z)$ become

\begin{equation}
Ec_{\beta }(z)|_{\beta =1}=Ec_{1}(z)=\sum\limits_{m=0}^{\infty }\frac{%
(-1)^{m}z^{2m}}{\Gamma (2m+1)}=\sum\limits_{m=0}^{\infty }\frac{%
(-1)^{m}z^{2m}}{(2m)!}=\cos (z),  \label{eq14_c}
\end{equation}

and

\begin{equation}
Es_{\beta }(z)|_{\beta =1}=Es_{1}(z)=\sum\limits_{m=0}^{\infty }\frac{%
(-1)^{m}z^{(2m+1)}}{\Gamma ((2m+1)+1)}=\sum\limits_{m=0}^{\infty }\frac{%
(-1)^{m}z^{(2m+1)}}{(2m+1)!}=\sin (z).  \label{eq14_s}
\end{equation}

The new expression for the Mittag-Leffler function $E_{\beta }(i^{\beta }z)$
given by Eq.(\ref{eq57}) can be considered as a fractional generalization of
the celebrated Euler equation, which is recovered from Eq.(\ref{eq57}) in
the limit case $\beta =1$,

\begin{equation}
e^{iz}=\cos (z)+i\sin (z).  \label{eq16_sol}
\end{equation}

Two new expressions (\ref{eq15_sol1new}) and (\ref{eq15_sol2new}) can be
considered as a fractional generalization of the well-known equations

\begin{equation}
\cos (z)=\frac{1}{2}(e^{iz}+e^{-iz}),  \label{eq17_sol}
\end{equation}

and

\begin{equation}
\sin (z)=\frac{1}{2i}(e^{iz}-e^{-iz}).  \label{eq18_sol}
\end{equation}

Let us note, that Eq.(\ref{eq57}) can be further generalized to

\begin{equation}
E_{\beta ,\gamma }(i^{\sigma }z)=Ec_{\beta ,\gamma }^{\sigma }(z)+i^{\sigma
}Es_{\beta ,\gamma }^{\sigma }(z),  \label{eq19_sol}
\end{equation}

\begin{equation}
0<\beta \leq 1,\qquad 0<\gamma \leq 1,\qquad 0<\sigma \leq 1,
\end{equation}

where function $E_{\beta ,\gamma }(z)$ is defined by the series

\begin{equation}
E_{\beta ,\gamma }(z)=\sum\limits_{m=0}^{\infty }\frac{z^{m}}{\Gamma (\beta
m+\gamma )},
\end{equation}

and two new functions $Ec_{\beta ,\gamma }^{\sigma }(z)$ and $Es_{\beta
,\gamma }^{\sigma }(z)$ are introduced by

\begin{equation}
Ec_{\beta ,\gamma }^{\sigma }(z)=\sum\limits_{m=0}^{\infty }\frac{%
(-1)^{m\sigma }z^{2m}}{\Gamma (2\beta m+\gamma )},
\end{equation}

and

\begin{equation}
Es_{\beta ,\gamma }^{\sigma }(z)=\sum\limits_{m=0}^{\infty }\frac{%
(-1)^{m\sigma }z^{(2m+1)}}{\Gamma ((2m+1)\beta +\gamma )}.
\end{equation}

respectively.

In terms of functions $Ec_{\beta ,\gamma }^{\sigma }(z)$ and $Es_{\beta
,\gamma }^{\sigma }(z)$, the functions $Ec_{\beta }(z)$ and $Es_{\beta }(z)$
defined by Eqs.(\ref{eq13_sol}) and (\ref{eq14_sol}) are given by

\begin{equation}
Ec_{\beta }(z)=Ec_{\beta ,\gamma }^{\sigma }(z)|_{\sigma =\beta ,\gamma
=1},\qquad Es_{\beta }(z)=Es_{\beta ,\gamma }^{\sigma }(z)|_{\sigma =\beta
,\gamma =1}.
\end{equation}

\section{Appendix B}

\subsection{Fox's $H$-function}

The $H$-function introduced by Fox \cite{Fox}, possesses many interesting
properties which can be used in time fractional quantum mechanics to
calculate integrals and perform transformations to study limiting cases at
particular choices of fractality parameters.

Fox's $H$-function is defined by the Mellin-Barnes type integral \cite{Fox}, 
\cite{Braaksma}, \cite{Mathai1} (we follow the notations of the book \cite%
{Mathai1})

\begin{equation}
H_{p,q}^{m,n}(z)=H_{p,q}^{m,n}\left[ z|\QATOP{(a_{p},A_{p})}{(b_{q},B_{q})}%
\right]  \label{eqA.1}
\end{equation}

\begin{equation*}
=H_{p,q}^{m,n}\left[ z\mid \QATOP{(a_{1},A_{1}),...,(a_{p},A_{p})}{%
(b_{1},B_{1}),...,(b_{q},B_{q})}\right] =\frac{1}{2\pi i}\int%
\limits_{C}ds~z^{s}~\chi (s),
\end{equation*}

where function $\chi (s)$ is given by

\begin{equation}
\chi (s)=\frac{\prod\limits_{j=1}^m\Gamma
(b_j-B_js)\prod\limits_{j=1}^n\Gamma (1-a_j+A_js)}{\prod\limits_{j=m+1}^q%
\Gamma (1-b_j+B_js)\prod\limits_{j=n+1}^p\Gamma (a_j-A_js)},  \label{eqA.2}
\end{equation}

and

\begin{equation*}
z^{s}=\exp \{s\mathrm{Log}|z|+i\arg z\},
\end{equation*}%
here $m$,  $n$,  $p$ and $q$ are non negative integers satisfying $0\leq
n\leq p$, 1$\leq m\leq q$; and the empty products are interpreted as unity.
The parameters $A_{j}$ ($j=1,...,p$) and $B_{j}$ ($j=1,...,q$) are positive
numbers; $a_{j}$ ($j=1,...,p$) and $b_{j}$ ($j=1,...,p$) are complex numbers
such that

\begin{equation}
A_j(b_h+\nu )\neq B_h(a_j-\lambda -1)  \label{eqA.3}
\end{equation}
for $\nu $, $\lambda =0,1,...$; $h=1,...,m$; $j=1,...,n$.

The $C$ is a contour separating the points

\begin{equation*}
s=\left( \frac{b_j+\nu }{B_j}\right) ,\quad (j=1,...,m;~~\nu =0,1,...),
\end{equation*}

which are the poles of $\Gamma (b_j-B_js)$ $(j=1,...,m)$, from the points

\begin{equation*}
s=\left( \frac{a_j-\nu -1}{A_j}\right) ,\quad (j=1,...,n;~~\nu =0,1,...),
\end{equation*}

which are the poles of $\Gamma (1-a_{j}+A_{j}s)$ $(j=1,...,n)$. The contour $%
C$ exists on account of (\ref{eqA.3}). These assumptions will be retained
throughout.

In the contracted form the $H$-function in (\ref{eqA.1}) will be denoted by
one of the following notations:

\begin{equation}
H(z),\quad H_{p,q}^{m,n}(z),\quad H_{p,q}^{m,n}\left[ z|\QATOP{(a_{p},A_{p})%
}{(b_{q},B_{q})}\right] .  \label{eqA4}
\end{equation}

The Fox $H$-function is an analytic function of $z$ which makes sense (i)
for every $z\neq 0$ if $\mu >0$ and (ii) for $0<|z|<\beta ^{-1}$ if $\mu =0$%
, where

\begin{equation}
\mu =\sum\limits_{j=1}^{q}B_{j}-\sum\limits_{j=1}^{p}A_{j}  \label{eqA_mu}
\end{equation}

and

\begin{equation}
\beta
=\prod\limits_{j=1}^{p}A_{j}^{A_{j}}\prod\limits_{j=1}^{q}B_{j}^{-B_{j}}.
\label{eqA_beta}
\end{equation}

Due to the occurrence of the factor $z^{s}$ in (\ref{eqA.1}), the $H$
function is in general multiple-valued, but is one-valued on the Riemann
surface of $\log z$.

The Mittag-Leffler $E_{\beta }(z)$ function \cite{Mittag-Leffler} given by
series (\ref{eq52}) has the following representation in terms of the $H$%
-function

\begin{equation}
E_{\beta }(z)=H_{1,2}^{1,1}\left[ -z|\QATOP{(0,1)}{(0,1),(0,\beta )}\right]
\qquad 0<\beta \leq 1.  \label{eqA5}
\end{equation}

To represent an $H$-function in computable form let us consider the case
when the poles $s=(b_{j}+\nu )/B_{j}$ ($j=1,...,m;$ $\nu =0,1,...$) of $%
\prod\limits_{j=1}^{m}{}^{^{\prime }}\Gamma (b_{j}-B_{j}s)$ are simple, that
is, where

\begin{equation*}
B_h(b_j+\lambda )\neq B_j(b_h+\nu ),\qquad j\neq h,
\end{equation*}

\begin{equation*}
h=1,...,m;\quad \quad \quad \quad \nu ,\lambda =0,1,2,...,
\end{equation*}

and the prime means the product without the factor $j=h$.

Then we obtain the following expansion for the $H$-function

\begin{equation}
H_{p,q}^{m,n}(z)=\sum\limits_{h=1}^m\sum\limits_{k=0}^\infty \frac{%
\prod\limits_{j=1}^n\Gamma (1-a_j+A_js_{hk})\prod\limits_{j=1}^m{}^{^{\prime
}}\Gamma (b_j-B_js_{hk})}{\prod\limits_{j=m+1}^q\Gamma
(1-b_j+B_js_{hk})\prod\limits_{j=n+1}^p\Gamma (a_j-A_js_{hk})}\frac{(-1)^k}{%
k!}\frac{z^{s_{hk}}}{B_h},  \label{eqA.4}
\end{equation}

\begin{equation*}
s_{hk}=(b_h+k)/B_h
\end{equation*}

which exists for all $z\neq 0$ if $\mu >0$ and for $0<|z|,\beta ^{-1}$ if $%
\mu =0$, where $\mu $ and $\beta $ are given by the Eqs.(\ref{eqA_mu}) and (%
\ref{eqA_beta}). The prime means the product without the factor $j=h$.

The formula (\ref{eqA.4}) can be used for the calculation of special values
of the Fox function and to derive the asymptotic behavior for $z\rightarrow
0 $.

The asymptotic expansion for $|z|\rightarrow \infty $ are treated in Ref.%
\cite{Braaksma} in the general case. In particular, for $\mu >0$ and $n\neq
0 $

\begin{equation*}
H_{p,q}^{m,n}(z)\sim \sum \mathrm{res}(\chi (s)~z^s),
\end{equation*}

as $|z|\rightarrow \infty $ uniformly on every closed subsector of $|\arg
z|\leq \frac 12\pi \lambda $. The residues have to be taken at the points $%
s=(a_j-1-\nu )/\alpha _j$ ($j=1,...,n$; $\nu =0,1,...$) and $\lambda $ is
defined by

\begin{equation}
\lambda
=\sum\limits_{j=1}^{m}B_{j}+\sum\limits_{j=1}^{n}A_{j}-\sum%
\limits_{j=m+1}^{q}B_{j}-\sum\limits_{j=n+1}^{p}A_{j}.  \label{eqA4_1}
\end{equation}

Symmetries in the parameters of the $H$ function are detected by regarding
the definitions (\ref{eqA.1}) and (\ref{eqA.2}).

\subsubsection{Some properties of the $H$-function}

Here we present a list of mainly used properties of the $H-$function. The
results of this section follow readily from the definition of the $H-$%
function (\ref{eqA.1}) and hence no proofs are given here.

\textbf{Property 1} The $H-$function is symmetric in the pairs $%
(a_{1},A_{1}),...,(a_{n},A_{n})$, likewise $%
(a_{n+1},A_{n+1}),...,(a_{p},A_{p})$; in $(b_{1},B_{1}),...,(b_{m},B_{m})$
and in $(b_{m+1},B_{m+1}),...,(b_{q},B_{q})$.

\textbf{Property 2} If one of the $(a_{j},A_{j})$ $(j=1,...,n)$ is equal to
one of the $(b_{j},B_{j})$ $(j=m+1,...,q)$ or one of the $(b_{j},B_{j})$ $%
(j=1,...,m)$ is equal to one of the $(a_{j},A_{j})$ $(j=n+1,...,p)$, then
the $H-$function reduces to one of the lower order, and $p,q$ and $n$ (or $%
m) $ decrease by unity.

Thus, we have the following reduction formulas,

\begin{equation}
H_{p,q}^{m,n}z\mid \left[ \QATOP{(a_{1},A_{1}),...,(a_{p},A_{p})}{%
(b_{1},B_{1}),...,(b_{q-1},B_{q-1}),(a_{1},A_{1})}\right]  \label{eqA.5}
\end{equation}

\begin{equation*}
=H_{p-1,q-1}^{m,n-1}\left[ z\mid \QATOP{(a_{2},A_{2}),...,(a_{p},A_{p})}{%
(b_{1},B_{1}),...,(b_{q-1},B_{q-1})}\right] ,
\end{equation*}

provided $n\geq 1$ and $q>m$, \ and

\begin{equation}
H_{p,q}^{m,n}\left[ z\mid \QATOP{%
(a_{1},A_{1}),...,(a_{p-1},A_{p-1}),(b_{1},B_{1})}{%
(b_{1},B_{1}),...,(b_{q-1},B_{q-1}),(b_{q},B_{q})}\right]  \label{eqA_6}
\end{equation}

\begin{equation*}
=H_{p-1,q-1}^{m-1,n}\left[ z\mid \QATOP{(a_{1},A_{1}),...,(a_{p-1},A_{p-1})}{%
(b_{2},B_{2}),...,(b_{q},B_{q})}\right] ,
\end{equation*}

provided $m\geq 1$ and $p>n$.

\textbf{Property 3}

\begin{equation}
H_{p,q}^{m,n}\left( z\mid \QATOP{(a_{1},A_{1}),...,(a_{p},A_{p})}{%
(b_{1},B_{1}),...,(b_{q},B_{q})}\right)  \label{eqA.6}
\end{equation}

\begin{equation*}
=H_{q,p}^{n,m}\left( \frac{1}{z}\mid \QATOP{%
(1-b_{1},B_{1}),...,(1-b_{q},B_{q})}{(1-a_{1},A_{1}),...,(1-a_{p},A_{p})}%
\right) ,
\end{equation*}

This is an important property of the $H$-function because it enables us to
transform an $H-$function with $\mu
=\sum\limits_{j=1}^{m}B_{j}-\sum\limits_{j=1}^{n}A_{j}>0$ and $\arg x$ to
one with $\mu <0$ and $\arg (1/x)$ and vice versa.

\textbf{Property 4}

\begin{equation}
\frac{1}{k}H_{p,q}^{m,n}\left( z\mid \QATOP{(a_{1},A_{1}),...,(a_{p},A_{p})}{%
(b_{1},B_{1}),...,(b_{q},B_{q})}\right) =H_{p,q}^{m,n}\left( z^{k}\mid 
\QATOP{(a_{1},kA_{1}),...,(a_{p},kA_{p})}{(b_{1},kB_{1}),...,(b_{q},kB_{q})}%
\right) ,  \label{eqA.7}
\end{equation}

where $k>0$.

\textbf{Property 5}

\begin{equation}
z^{\sigma }H_{p,q}^{m,n}\left( z\mid \QATOP{(a_{1},A_{1}),...,(a_{p},A_{p})}{%
(b_{1},B_{1}),...,(b_{q},B_{q})}\right)  \label{eqA.8}
\end{equation}

\begin{equation*}
=H_{p,q}^{m,n}\left( z\mid \QATOP{(a_{1}+\sigma
A_{1},A_{1}),...,(a_{p}+\sigma A_{p},A_{p})}{(b_{1}+\sigma
B_{1},B_{1}),...,(b_{q}+\sigma B_{q},B_{q})}\right) ,
\end{equation*}

Further interesting and important properties of the Fox $H$-function and
expression for elementary and special functions in terms of the $H$-function
are listed in Ref.\cite{Mathai1}.

\subsection{Cosine transform of the $H$-function}

The cosine transform of $H$-function $H_{p,q}^{m,n}\left[ at^{\beta }{\LARGE %
|}\QATOP{(a_{p},A_{p})}{(b_{q},B_{q})}\right] $ is defined by \cite%
{Prudnikov}

\begin{equation}
\int\limits_{0}^{\infty }dtt^{\nu -1}\cos (xt)H_{p,q}^{m,n}\left[ at^{\alpha
}{\LARGE |}\QATOP{(a_{p},A_{p})}{(b_{q},B_{q})}\right]  \label{eqB_1a}
\end{equation}%
\begin{equation*}
=\frac{\pi }{x^{\nu }}H_{q+1,p+2}^{m+1,n}\left[ \frac{x^{\beta }}{a}{\LARGE |%
}\QATOP{(1-b_{q},B_{q}),(\frac{\nu +1}{2},\frac{\alpha }{2})}{(\nu ,\alpha
),(1-a_{p},A_{p}),(\frac{\nu +1}{2},\frac{\alpha }{2})}\right] ,
\end{equation*}

where \textrm{Re}[$\nu +\alpha \min_{1\leq j\leq m}(\frac{b_{j}}{B_{j}})]>0,$
\textrm{Re}[$\nu +\alpha \max_{1\leq j\leq n}(\frac{a_{j}-1}{A_{j}})]<0$, $%
|\arg a|<\pi \lambda /2$, with $\lambda $ defined by Eq.(\ref{eqA4_1}).

\end{document}